\documentclass[%
 reprint,
superscriptaddress,
 amsmath,amssymb,
 aps,
]{revtex4-2}

\usepackage{ulem}
\usepackage{physics}
\usepackage{graphicx}
\usepackage{bm}
\usepackage{booktabs}
\usepackage{color}
\usepackage{hyperref}

\begin{document}


\title{Self-organized photonic time quasicrystal from a single imposed clock}

\author{Minwook Kyung}
\thanks{These authors contributed equally to this work.}
\affiliation{%
 Department of Physics, Korea Advanced Institute of Science and Technology, Daejeon 34141, Republic of Korea
}%
\author{Kyungmin Lee}
\thanks{These authors contributed equally to this work.}
\affiliation{%
 Department of Physics, Korea Advanced Institute of Science and Technology, Daejeon 34141, Republic of Korea
}
\author{Yung Kim}
\affiliation{%
 Department of Physics, Korea Advanced Institute of Science and Technology, Daejeon 34141, Republic of Korea
}%
\author{Eun-Gook Moon}
\affiliation{%
 Department of Physics, Korea Advanced Institute of Science and Technology, Daejeon 34141, Republic of Korea
}
\author{Joonhee Choi}
\affiliation{%
 Department of Electrical Engineering, Stanford University, Stanford, CA 94305, USA
}
\author{Bumki Min}
 \email{bmin@kaist.ac.kr}
\affiliation{%
 Department of Physics, Korea Advanced Institute of Science and Technology, Daejeon 34141, Republic of Korea
}

\date{\today}

\begin{abstract}
A photonic time crystal usually writes a clock into a medium. We show that a nonlinear medium driven by one such clock selects a quasiperiodic temporal order for itself. In a guided-wave lattice of nonlinear dipoles, a single-tone pump modulates the polarization sector, while Maxwell--polarization back-action selects two response frequencies tied to the drive by one resolved independent low-order relation, the pump-locked sum condition. Their sum phase locks to the pump and the complementary phase winds, producing a photonic discrete time quasicrystal with torus-like phase dynamics and a discrete combination spectrum. Simultaneous site-resolved measurements show that both the locked sum phase and the winding complementary phase are spatially coherent, and the two-frequency response spans a range of drive power and injected noise. Tuned to a degenerate resonance, the same pump drives the lattice into a period-doubled time crystal, so the frequency of a single tone selects between periodic and quasiperiodic self-organized order in one medium.
\end{abstract}

\maketitle

A photonic time crystal writes a clock into a medium. In nearly all realizations, that clock is imposed from outside through a time-periodic constitutive response, which couples Floquet harmonics separated by the modulation frequency and, in a spatially uniform medium, hybridizes frequencies while preserving momentum; this Maxwell--Floquet coupling opens momentum gaps~\cite{reyes2015observation,PhysRevA.93.063813,asgari2024theory,wang2025expanding,PhysRevB.98.085142} and enables pairwise amplification of counter-propagating waves, frequency conversion, and modified emission and absorption pathways~\cite{galiffi2022photonics,park2025spontaneous,lee2026analogs,sustaeta2025quantum,bae2025quantum,dong2025extremely,feinberg2025plasmonic,allard2026broadband}. In these realizations the temporal modulation is treated as an externally prescribed input, independent of the evolving electromagnetic state~\cite{PhysRevA.93.063813,park2022revealing,lyubarov2022amplified}, and this description has been effective across microwave, metasurface, and parametric implementations~\cite{wang2023metasurface,martinez2018parametric}. The modulated elements of such platforms are often resonant and nonlinear, so the same media can host a response beyond the programmed clock. Once the driven polarization coordinate is nonlinear and acts back on the electromagnetic field, the clock and the response are no longer separate. In such a Maxwell--polarization medium, the imposed clock fixes only the external discrete time-translation symmetry, and the long-time state can select additional phases that the modulation waveform does not specify.

In periodically driven many-body systems, discrete time-translation symmetry can be broken so that long-time observables recur with period \(qT\), an integer multiple \(q>1\) of the drive period \(T\)~\cite{PhysRevLett.109.160401,PhysRevLett.109.160402,PhysRevLett.117.090402,PhysRevLett.118.030401,else2020discrete,yi2024theory}. Such a time-crystalline response is selected by the dynamics, whereas a merely driven response follows the drive. This principle has been observed in quantum platforms~\cite{choi2017observation,zhang2017observation,kyprianidis2021observation}, and related dissipative subharmonic order has been explored in optical resonators~\cite{taheri2022all}. Its defining signature is rigidity beyond frequency division: a response that persists over a finite parameter range against perturbations that preserve the drive periodicity~\cite{yao2020classical,PhysRevLett.123.124301,zaletel2023colloquium}. Photonics also offers a continuous-time counterpart, in which time-independent coherent illumination drives plasmonic metamaterials into persistent, spatially ordered oscillations, including continuous space--time crystal states driven by nonreciprocal optical forces~\cite{liu2023photonic,raskatla2024continuous}. For a Floquet photonic medium, the corresponding question is whether one imposed modulation period can generate an additional temporal phase degree of freedom through the internal dynamics of a light--matter medium.

Quasiperiodicity is a demanding test of this self-selection, because a second incommensurate imposed tone would directly provide an independent frequency. In the sum-resonant channel investigated here, quasiperiodic self-selection appears as two response frequencies tied to the drive by one resolved independent low-order relation, the pump-locked sum condition. For two resonant modes pumped above threshold at the sum of their frequencies, the pump constrains the sum of the two oscillation phases and leaves their difference free~\cite{bengtsson2018nondegenerate}, and anharmonic responses of periodically driven arrays through the coresonance of two dispersion branches have been predicted~\cite{nicolaou2021anharmonic}. We ask whether a photonic medium shares this free phase across its sites, so that the quasiperiodic response becomes a collective state. Spatial quasicrystals established that long-range order need not rely on a primitive lattice period~\cite{shechtman1984metallic,levine1984quasicrystals}, and recent work has extended this principle to time-quasicrystalline phases~\cite{autti2018observation,PhysRevLett.123.150601,PhysRevB.100.134302,PhysRevX.10.021032,nicolaou2021anharmonic,PhysRevX.15.011055,feng2025tunable}. In a Floquet photonic medium, the corresponding state is a two-frequency temporal order in which a pump-locked sum phase remains rigid while a complementary phase winds. When the winding is incommensurate with the drive, the stroboscopic section of such a state is an invariant circle, that is, the motion lies on a torus.

\begin{figure*}[htb!]
  \centering
\includegraphics[width=0.75\textwidth]{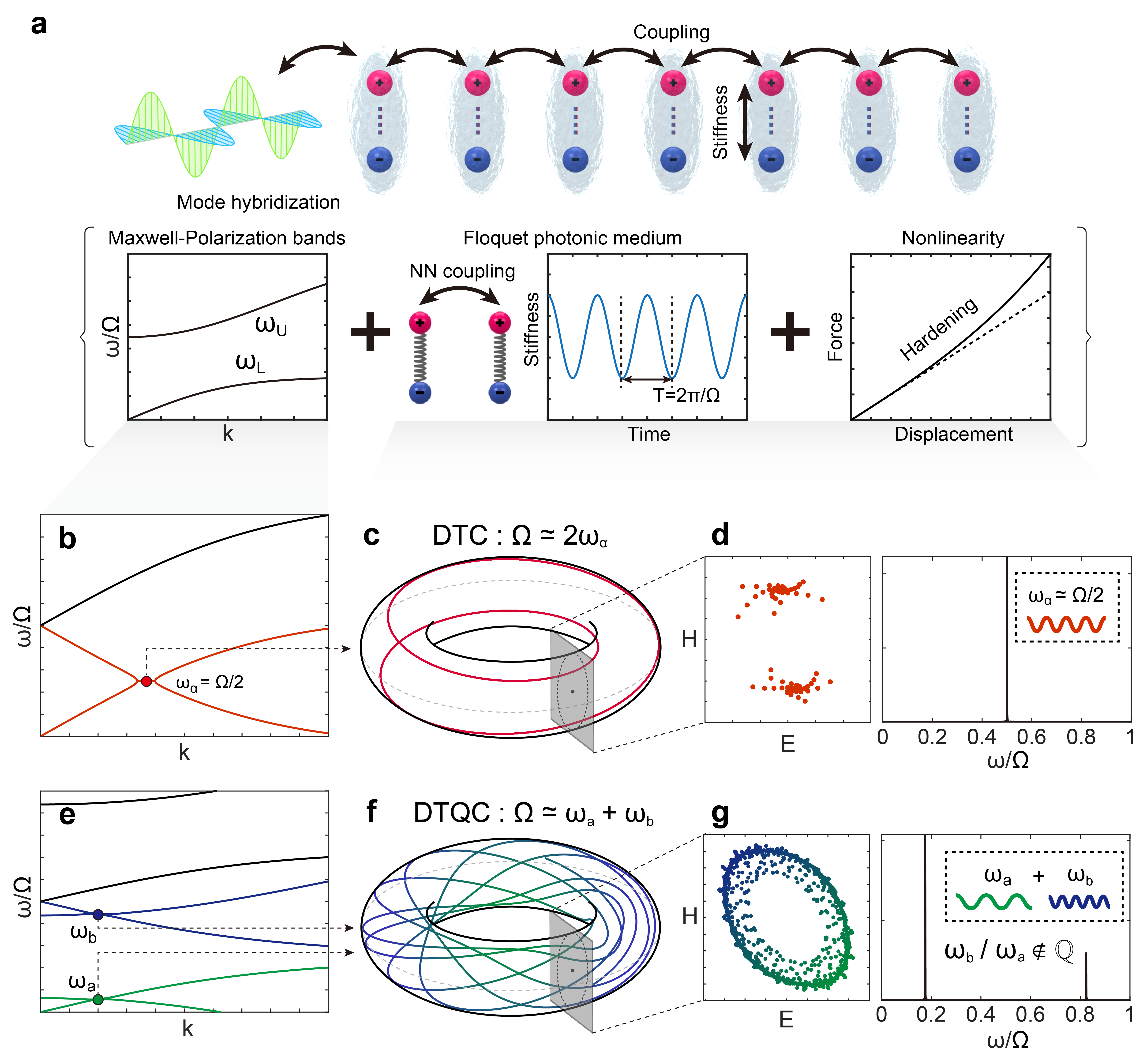}
    \caption{\label{fig:Fig1}
    Single-clock mechanism for self-organized temporal order in a Maxwell--polarization lattice.
    (a) Guided electromagnetic mode hybridized with a Lorentz-type nonlinear polarization resonance, with nearest-neighbour polarization coupling, stiffness modulation at angular frequency \(\Omega\), and on-site Duffing nonlinearity.
    (b--d) DTC sector, \(\Omega\simeq2\omega_\alpha\): period-doubled limit cycle, two-point stroboscopic section, and subharmonic spectral peak.
    (e--g) DTQC sector, \(\Omega\simeq\omega_a+\omega_b\) with carriers from the lower and upper hybrid bands: invariant torus and a spectrum with two fundamental response frequencies.
}
\end{figure*}

Here we experimentally demonstrate this photonic counterpart in a microwave guided-wave lattice of varactor-loaded nonlinear LC resonators. The resonators provide pumped nonlinear polarization coordinates, and the guided electromagnetic mode hybridizes with them and mediates coupling across the array. Field-resolved access to the array lets us resolve the ingredient usually hidden in optical time-varying media, namely the self-consistent back-action between the electromagnetic field and the driven polarization sector. We identify the resulting spatially correlated two-frequency response as a photonic discrete time quasicrystal (DTQC), a quasiperiodic Floquet state selected by the medium rather than programmed into the drive.

Four properties define the photonic DTQC. The drive has a single period \(T=2\pi/\Omega\), and the steady state has two internally selected spectral fundamentals. Their sum phase is locked to the pump, and the complementary phase winds, with the sum condition as the one independent low-order relation resolved in a scan of frequency combinations. Both phases are coherent across the simultaneously recorded lattice sites, and the two-frequency response spans a range of drive power and noise.

We start from a continuum Maxwell--Lorentz description of a guided-wave medium with a resonant nonlinear polarization coordinate \(P(z,t)\). The electric and magnetic fields \(E\) and \(H\) and the polarization obey
\begin{align}
\partial_z E&=-\mu\,\partial_t H,\qquad
\partial_z H=-\partial_t(\varepsilon_bE+P),
\label{eq:maxwell_closure}\\
\ddot P&+\gamma\dot P+\omega_P^2(t)P+\beta P^3
-v_P^2\partial_z^2P=\chi E ,
\label{eq:continuum_model}
\end{align}
where \(\mu\) and \(\varepsilon_b\) are the permeability and background permittivity of the guide, \(\gamma\) is the polarization damping rate, \(\beta\) the local Duffing nonlinearity, \(v_P\) the polarization-coupling velocity, and \(\chi\) the field--polarization coupling. The pump of frequency \(\Omega=2\pi/T\) modulates the polarization stiffness about \(\omega_0^2\), \(\omega_P^2(t)=\omega_0^2[1+\delta\cos(\Omega t)]\), so the imposed periodicity enters through an internal material degree of freedom, in place of a prescribed \(\varepsilon(t)\) inserted directly into Maxwell's equations. Here \(\delta\) denotes the effective stiffness-modulation depth of the Maxwell--polarization model, and its relation to the circuit-level resonator-capacitance modulation is given in Supplementary Section~S2.

In optical and polaritonic platforms, \(P\) may represent an electronic, excitonic, phononic or polaritonic polarization coordinate, while \(\omega_P(t)\) or the corresponding oscillator strength can be tuned by optical or electrical pumping~\cite{Shcherbakov2019,wang2025expanding,allard2026broadband}. The cubic term is a Duffing nonlinearity of this polarization oscillator, and after projection onto the hybrid Maxwell--polarization branches it produces Kerr-like self- and cross-nonlinearities. The underlying response nevertheless remains resonant and dynamical, distinct from an instantaneous Kerr constitutive law. In the experimental microwave resonator array, \(n=1,\ldots,N\) labels the unit cell, and the continuum coupling term is implemented by nearest-neighbour polarization coupling, 
\(-v_P^2\partial_z^2P\rightarrow g(2P_n-P_{n+1}-P_{n-1})\), with \(g\simeq v_P^2/d^2\) for unit-cell spacing \(d\). The circuit realization retains the guided-wave and dynamical polarization degrees of freedom of the Maxwell--Lorentz description, with the circuit-specific inductive network incorporated in Supplementary Sections~S2 and S3.

Experimentally, this medium is a one-dimensional guided-wave array of \(N=12\) varactor-loaded split-ring LC resonators (Supplementary Section~S4 and Supplementary Fig.~\ref{fig:Figsm}), in which a dc bias sets the varactor operating point and a common single-tone radio-frequency pump modulates the resonator capacitance and thereby the effective polarization stiffness. We access the DTC and DTQC sectors at modulation frequencies \(f_{\mathrm{mod}}=2.98~\mathrm{GHz}\) and \(2.47~\mathrm{GHz}\), respectively. Site-resolved voltages are recorded simultaneously, while calibrated broadband noise is injected from one end of the waveguide to tune \(S_{\mathrm{noise}}\).

In the unmodulated linear limit, the Maxwell--Lorentz model of Fig.~\ref{fig:Fig1}(a) supports two polariton-like hybrid bands, the lower and upper branches \(\omega_{\mathrm L}(k)\) and \(\omega_{\mathrm U}(k)\) (Supplementary Section~S1), and periodic stiffness modulation selects near-resonant carrier frequencies from these bands. In a degenerate channel, a selected carrier satisfies \(\Omega\simeq2\omega_{\alpha}\) and produces a local period-doubled response (Fig.~\ref{fig:Fig1}(b--d)), whereas in a nondegenerate channel two selected carriers, \(\omega_a\simeq\omega_{\mathrm L}(k_0)\) and \(\omega_b\simeq\omega_{\mathrm U}(k_0)\), with \(k_0\) denoting the resonant momentum window, satisfy the sum-resonance condition \(\Omega\simeq\omega_a+\omega_b\), so that the pump couples their slow sum phase (Fig.~\ref{fig:Fig1}(e--g)). The degenerate case thus gives the local DTC channel and the nondegenerate case the local DTQC channel. In the latter, the common pump locks the local sum phase, while electromagnetic coupling synchronizes the winding complementary phase across the array and establishes collective quasiperiodic order.

\begin{figure*}[htb!]
  \centering
    \includegraphics[width=1.0\textwidth]{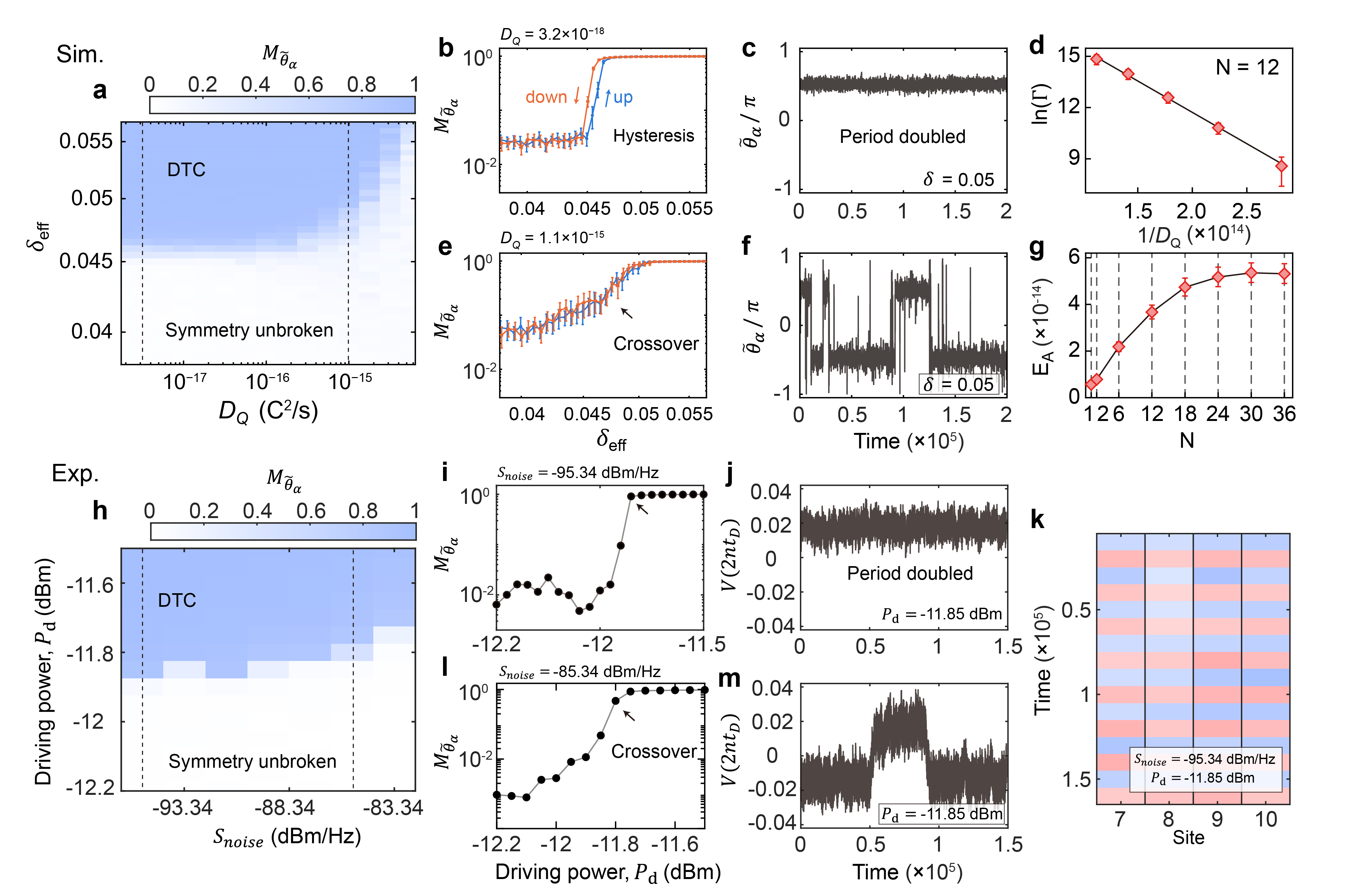}
   \caption{\label{fig:Fig3}
    Collective DTC order in the degenerate reference sector. (a--g) Simulation. (a) Phase memory \(M\) in the \((D_Q,\delta_{\rm eff})\) plane. (b,c) \(M\) versus \(\delta_{\rm eff}\) for up and down sweeps and rotating-frame phase versus time at lower charge noise. (d) Logarithm of the escape rate, \(\ln\Gamma\), versus \(1/D_Q\) (in \(\mathrm{s/C^2}\)) for \(N=12\). (e,f) Same as (b,c) at stronger noise. (g) Activation coefficient \(E_A\) (in \(\mathrm{C^2/s}\)) versus array size \(N\). (h--m) Experiment. (h) Measured \(M\) in the $(S_{\mathrm{noise}},P_d)$ plane. (i--k) \(M\) versus \(P_d\), voltage trace, and site--time voltage map over the recorded sites at $S_{\mathrm{noise}}=-95.34\,\mathrm{dBm/Hz}$. (l,m) \(M\) versus \(P_d\) and voltage trace at $S_{\mathrm{noise}}=-85.34\,\mathrm{dBm/Hz}$.}
\end{figure*}

\begin{figure*}[htb!]
  \centering
  \includegraphics[width=1.0\textwidth]{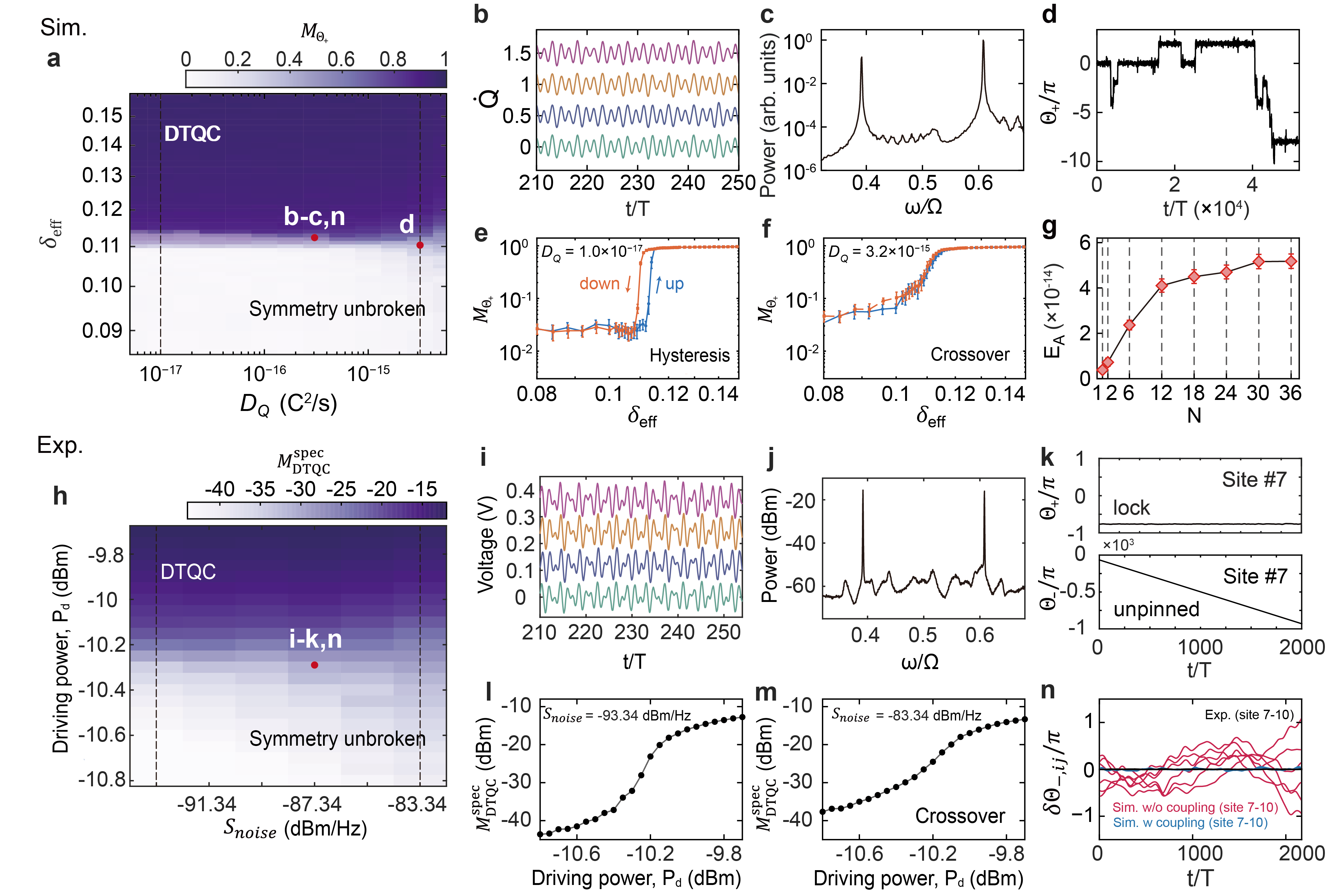}
  \caption{\label{fig:Fig4} Collective DTQC order in the nondegenerate sum-resonant sector. (a--g) Simulation. (a) Sum-phase memory \(M_{\Theta_+}\) in the \((D_Q,\delta_{\rm eff})\) plane. (b,c) Resonator charge currents \(\dot Q_{r,n}\) at the sites recorded in the experiment (vertically offset) and site-averaged power spectrum at the point marked b--c in (a). (d) Lifted sum phase \(\Theta_+/\pi\) versus time at the point marked d in (a). (e,f) \(M_{\Theta_+}\) versus \(\delta_{\rm eff}\) for up and down sweeps at \(D_Q=1.0\times10^{-17}\) and \(3.2\times10^{-15}\,\mathrm{C^2/s}\), respectively. (g) Activation coefficient \(E_A\) (in \(\mathrm{C^2/s}\)) versus array size \(N\). (h--m) Experiment. (h) Two-line spectral weight \(M_{\mathrm{DTQC}}^{\mathrm{spec}}\) in the \((S_{\mathrm{noise}},P_d)\) plane. (i,j) Voltages at the recorded sites and spectrum at the selected operating point marked in (h). (k) Sum phase \(\Theta_+\) and complementary phase \(\Theta_-\) versus time at one recorded site. (l,m) Two-line spectral weight versus \(P_d\) at \(S_{\mathrm{noise}}=-93.34\) and \(-83.34\,\mathrm{dBm/Hz}\). (n) Mean-centered relative complementary phases \(\delta\Theta_{-,ij}(t)/\pi\) versus time across the recorded sites from the experiment and from simulations with and without inter-site coupling.}
\end{figure*}

The two locking channels follow from a branch-resolved normal form, in which we project the driven Maxwell--polarization dynamics onto the selected near-resonant carrier sectors and retain the slow action--angle variables \(J_{\alpha,n}\) and \(\tilde{\theta}_{\alpha,n}\), where \(\alpha=a,b\) labels the selected carrier sector and \(n\) labels the unit cell. The coefficients below are effective slow-Hamiltonian parameters obtained after hybrid-branch projection and rotating-wave averaging (Supplementary Section~S2). In the circuit dynamics, parametric pumping, nonlinear frequency shifts, and linear damping determine the steady oscillation amplitudes, while additive noise on the resonator charges, of intensity \(D_Q\), produces fluctuations and switching between attracting states. The conservative part of the leading-order stroboscopically averaged dynamics is the Hamiltonian
\begin{equation}
\label{eq:slow_hamiltonian}
\begin{aligned}
\bar{\mathcal{H}}
&=\sum_{n=1}^{N}\Big[
\sum_{\alpha}\Big\{
\Delta_\alpha J_{\alpha,n}
+\frac{\Lambda_{\alpha\alpha}}{2}J_{\alpha,n}^2
+\frac{h_\alpha}{2}J_{\alpha,n}
\cos 2\tilde{\theta}_{\alpha,n}
\Big\}
\\
&\;+\Lambda_{ab}J_{a,n}J_{b,n}
- g_{ab}\sqrt{J_{a,n}J_{b,n}}
\cos\!\bigl(\tilde{\theta}_{a,n}+\tilde{\theta}_{b,n}\bigr)
\Big]
\\
&\;-\sum_{n=1}^{N-1}\sum_{\alpha=a,b}
K_\alpha \sqrt{J_{\alpha,n}J_{\alpha,n+1}}\,
\cos\!\bigl(\tilde{\theta}_{\alpha,n}-\tilde{\theta}_{\alpha,n+1}\bigr),
\end{aligned}
\end{equation}
where \(\tilde{\theta}_{\alpha,n}\) are phases in the frame rotating at \(\Omega/2\), so that a branch oscillates at the laboratory frequency \(\Omega/2+\dot{\tilde{\theta}}_{\alpha,n}\), \(\Delta_\alpha\) are the on-site detunings from \(\Omega/2\), \(\Lambda_{\alpha\alpha}\) and \(\Lambda_{ab}\) are the Kerr-like self- and cross-nonlinearities, and \(K_\alpha\) is the nearest-neighbour hopping amplitude of branch \(\alpha\). Equation~\eqref{eq:slow_hamiltonian} lists the terms for interior sites, and the correction at the two open ends of the array, where a site has a single neighbour, is given in Supplementary Section~S3. The \(h_\alpha\) terms are intrabranch \(2{:}1\) locking terms, which pin \(2\tilde{\theta}_{\alpha,n}\) and generate the two symmetry-related subharmonic sectors of a DTC, and in the nondegenerate sum-resonant channel they are off resonant and are omitted from the branch-reduced Hamiltonian. The retained interbranch pump channel \(g_{ab}\) then depends only on the sum phase \(\Theta_{+,n}=\tilde{\theta}_{a,n}+\tilde{\theta}_{b,n}\), and \(\bar{\mathcal H}\) is invariant under the shift \((\tilde{\theta}_{a,n},\tilde{\theta}_{b,n})\rightarrow(\tilde{\theta}_{a,n}+\vartheta,\tilde{\theta}_{b,n}-\vartheta)\) applied at all sites. Hence the local complementary phase \(\Theta_{-,n}=\tilde{\theta}_{a,n}-\tilde{\theta}_{b,n}\) enters \(\bar{\mathcal H}\) only through inter-site differences in the \(K_\alpha\) terms, and its uniform mode has no restoring term. The winding of this free phase produces torus-like phase dynamics, which reduce to a limit cycle when an additional constraint fixes this phase or one branch is suppressed. Direct circuit integration yields a dominant subharmonic peak in the degenerate channel and two fundamental peaks with bounded \(\Theta_+\) in the nondegenerate channel (Supplementary Section~S3).

The phase dynamics follow from \(\dot{\tilde{\theta}}_{\alpha,n}=\partial\bar{\mathcal H}/\partial J_{\alpha,n}\) (Supplementary Section~S3). At interior sites of the uniform state selected in the \(N=12\) circuit model, with \(h_\alpha=0\) and fixed positive saturated actions \(\bar J_a,\bar J_b\), the sum and complementary phases obey
\begin{equation}
\begin{aligned}
\dot\Theta_\pm
&=
d_a\pm d_b
-\frac{g_{ab}}{2}\left(\rho\pm\rho^{-1}\right)\cos\Theta_+ ,\\
d_a&=\Delta_a-K_a+\Lambda_{aa}\bar J_a+\Lambda_{ab}\bar J_b,
\end{aligned}
\label{eq:dtqc_phase_reduction}
\end{equation}
where \(d_a\) is the detuning of branch \(a\) including its hopping and nonlinear shifts at the saturated actions, \(d_b\) follows from \(a\leftrightarrow b\), \(d_a+d_b\) is the effective sum detuning \(\Delta_+^{\rm eff}\) of Supplementary Section~S3, and \(\rho=(\bar J_b/\bar J_a)^{1/2}\). At fixed actions, the upper sign is an Adler equation~\cite{adler1946study} with sum-locked fixed points throughout \(|d_a+d_b|\le(|g_{ab}|/2)(\rho+\rho^{-1})\), and the damped dynamics determine their stability. The lower sign contains no \(\Theta_-\), so no term restores the complementary phase: at a sum-locked fixed point \(\dot\Theta_-\) is constant, and the laboratory-frame frequencies are
\begin{equation}
\omega_{1,2}
=
\frac{\Omega}{2}
\mp
\frac{\bar J_b d_b-\bar J_a d_a}{\bar J_a+\bar J_b},
\label{eq:dtqc_frequencies}
\end{equation}
where 1 and 2 label the lower- and upper-branch lines. At fixed \(\Omega\), the lock enforces \(\omega_1+\omega_2=\Omega\), so changes in the detunings or Kerr shifts move the two lines by equal and opposite amounts. For equal branch damping, the uniform deterministic reduction gives \(\bar J_a=\bar J_b\) and \(\omega_2-\omega_1=d_b-d_a\).

About the uniform sum-locked state, the hopping terms of Eq.~\eqref{eq:slow_hamiltonian} give the complementary phase the elastic energy
\begin{equation}
\bar{\mathcal H}_{\rm hop}
\simeq
\mathrm{const}
+\frac{K_a\bar J_a+K_b\bar J_b}{8}
\sum_{n=1}^{N-1}\left(\Theta_{-,n+1}-\Theta_{-,n}\right)^2 .
\label{eq:dtqc_stiffness}
\end{equation}
In the array, the frequency of each hybrid band increases with wavevector away from \(k=0\), which corresponds to positive hopping amplitudes, \(K_a,K_b>0\) (Supplementary Section~S3). Differences of the complementary phase between neighbouring sites thus carry a positive stiffness that favors alignment. Their common phase remains free, and relaxation toward the aligned state follows the damped slow flow (Supplementary Section~S3).

We first use the degenerate \(2{:}1\) channel as a periodic reference sector (Fig.~\ref{fig:Fig3}) to calibrate collective subharmonic locking in the same Maxwell--polarization lattice before we test the more demanding sum-resonant DTQC channel. In simulations, we characterize the ordered phase through the complex phase memory averaged over a set of sites \(\mathcal S\),
\begin{equation}
Z_{\varphi}^{(\mathcal S)}(t)
\equiv
\frac{1}{N_{\mathcal S}}
\sum_{n\in\mathcal S}
e^{i\Delta\varphi_n(t)},
\label{eq:complex_phase_memory}
\end{equation}
where \(\Delta\varphi_n(t)\equiv\varphi_n(t)-\varphi_n(t_{\rm ref})\), \(t_{\rm ref}\) is the first time point of the post-transient analysis window, and \(N_{\mathcal S}=|\mathcal S|\) is the number of analyzed sites. In the DTC sector, \(\varphi_n=\tilde\theta_{\alpha,n}\) is the rotating-frame phase of the resonant hybrid branch, and the late-time phase-memory order parameter is \(M_{\tilde\theta_\alpha}=\big|\langle Z_{\tilde\theta_\alpha}^{(\mathcal S)}(t)\rangle_t\big|\), which identifies the ordered region in the \((D_Q,\delta)\) plane (Fig.~\ref{fig:Fig3}(a)). Within this region, the phase remains bounded around one of the two period-doubled states. Low-noise line cuts show a hysteretic onset, whereas stronger noise broadens the onset into a crossover and increases stochastic switching (Fig.~\ref{fig:Fig3}(b--f)). Noise-activated escape rates follow \(\ln\Gamma_{\rm esc}=a_N-E_A(N)/D_Q\), where \(a_N\) is the rate offset, and the fitted activation coefficient \(E_A\) approaches a size-independent plateau for the larger arrays (Fig.~\ref{fig:Fig3}(g); the escape criterion is given in Supplementary Section~S3)~\cite{PhysRevLett.86.3942,PhysRevLett.99.060601}.

Experimentally, the reference sector shows the same noise-controlled phenomenology. As a function of drive power \(P_d\) and injected noise spectral density \(S_{\mathrm{noise}}\), the measured phase diagram reveals a DTC window (Fig.~\ref{fig:Fig3}(h)). At lower injected noise, increasing \(P_d\) produces an abrupt onset of stable \(2T\)-periodic voltage oscillations that are phase-aligned across the recorded sites (Fig.~\ref{fig:Fig3}(i--k)), whereas at higher injected noise the onset broadens into a smooth crossover (Fig.~\ref{fig:Fig3}(l,m)).

In the sum-resonant channel, the pump locks the branch-sum phase \(\Theta_+=\tilde\theta_a+\tilde\theta_b\), while the complementary phase \(\Theta_-=\tilde\theta_a-\tilde\theta_b\) continues to wind, and simulations show that inter-site coupling synchronizes its winding across the array. We locate the numerical DTQC regime in the \((D_Q,\delta)\) plane using the phase memory of Eq.~\eqref{eq:complex_phase_memory}. From the laboratory phases \(\phi_{1,n}(t)\) and \(\phi_{2,n}(t)\) of the two filtered voltage components, we reconstruct \(\Theta_{+,n}(t)=\phi_{1,n}(t)+\phi_{2,n}(t)-\Omega t\), which gives the temporal phase memory \(M_{\Theta_+}\), i.e., the persistence of the sum phase relative to the pump. Within the ordered DTQC window of the numerical phase diagram (Fig.~\ref{fig:Fig4}(a)), the coupled array exhibits spatially correlated quasiperiodic motion (Fig.~\ref{fig:Fig4}(b)), a steady-state spectrum with two resolved fundamentals (Fig.~\ref{fig:Fig4}(c)), and a sum phase with long-lived locked plateaus interrupted by noise-activated escape and relocking events (Fig.~\ref{fig:Fig4}(d)).

At low noise intensity \(D_Q\), the DTQC onset exhibits hysteresis associated with long stochastic residence times, whereas stronger noise enhances switching and broadens the onset into a smooth crossover (Fig.~\ref{fig:Fig4}(e,f)). Escape rates from the sum-locked state follow the same law, with an activation coefficient \(E_A\) that increases with the number of coupled cells and approaches a plateau for the largest simulated arrays (Fig.~\ref{fig:Fig4}(g)).

\begin{figure}[t!]
  \centering
  \includegraphics[width=0.48\textwidth]{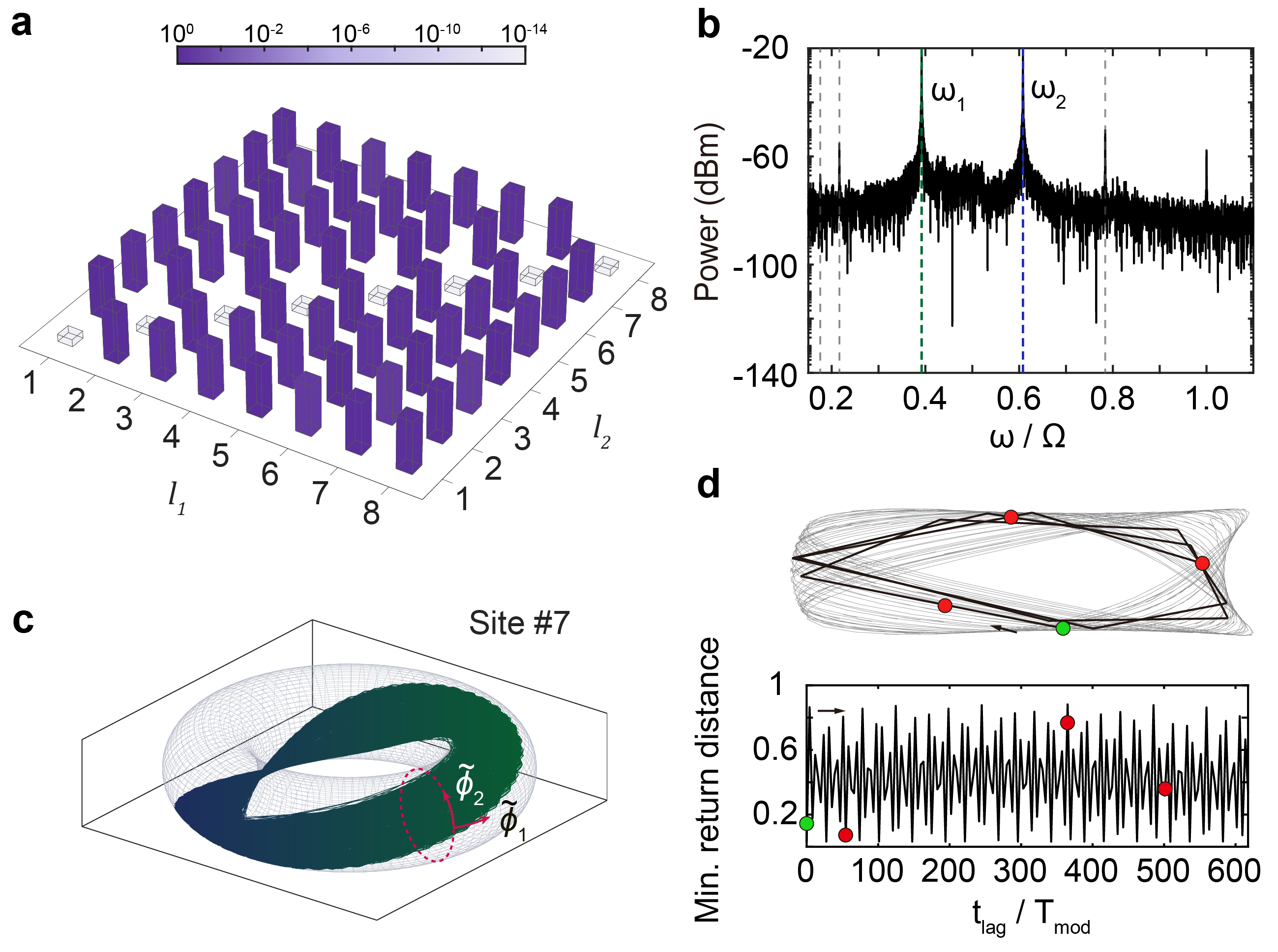}
  \caption{\label{fig:Fig5}
    Quasiperiodicity of the experimental DTQC state. (a) Normalized mismatch $\min_{l_3\in\mathbb Z} |l_1\omega_1+l_2\omega_2-l_3\Omega|/\Omega$ for integer pairs \((l_1,l_2)\) and the extracted response frequencies $(\omega_1,\omega_2)$. (b) Measured spectrum at the same operating point with combination lines of \(\omega_1\) and \(\omega_2\). (c) Two-phase trajectory at site \(n_0=7\) on the half-pump rotating phase torus \((\tilde\phi_{1,n_0},\tilde\phi_{2,n_0})\). (d) Normalized return distance of the phase trajectory at lags \(\tau=2mT\).}
\end{figure}

Experimentally, the two-line spectral weight $M_{\mathrm{DTQC}}^{\mathrm{spec}}=(A_1^{\mathrm{pk}}A_2^{\mathrm{pk}})^{1/2}$, where $A_j^{\mathrm{pk}}$ is the linear spectral peak power of the $j$-th response component, maps the two-frequency response window in the $(S_{\mathrm{noise}},P_d)$ plane (Fig.~\ref{fig:Fig4}(h)), within which phase reconstruction at a selected operating point, marked in Fig.~\ref{fig:Fig4}(h), establishes a sum-locked DTQC state. At this point, the voltages oscillate coherently across the sites (Fig.~\ref{fig:Fig4}(i)), the spectrum exhibits two clearly resolved peaks at \(\omega_1\) and \(\omega_2\) (Fig.~\ref{fig:Fig4}(j)), and the two independently extracted frequencies sum to \(\Omega\) with a fractional mismatch below $1.2\times10^{-6}$. Across the experimental parameter scan, the two frequencies shift in opposite directions while their sum stays constant within the frequency sampling of the measurement (Supplementary Section~S4). Phase reconstruction at this operating point, using response-frequency-resolved analytic signals at one recorded site, shows the locked sum phase and the winding complementary phase (Fig.~\ref{fig:Fig4}(k); Supplementary Section~S4). In the experimental line cuts at lower injected noise, \(S_{\mathrm{noise}}=-93.34~\mathrm{dBm/Hz}\), the two-line spectral weight shows a sharper onset as \(P_d\) is varied (Fig.~\ref{fig:Fig4}(l)), whereas at higher injected noise, \(S_{\mathrm{noise}}=-83.34~\mathrm{dBm/Hz}\), the onset broadens into a smooth crossover (Fig.~\ref{fig:Fig4}(m)).

Spatial order is measured by the pairwise relative-phase coherence of the reconstructed phases, with \(\Theta_{-,n}(t)=\phi_{1,n}(t)-\phi_{2,n}(t)\),
\begin{equation}
C_{ij}^{\pm}=\left|\left\langle e^{i[\Theta_{\pm,i}(t)-\Theta_{\pm,j}(t)]}\right\rangle_t\right| ,
\label{eq:Cij_main}
\end{equation}
which equals unity when the relative phase of two sites is constant in time. At this operating point, all six pairs of the four recorded sites give \(C_{ij}^{\pm}\approx1\) over the post-transient window (Supplementary Fig.~\ref{fig:Figsm_spatial_coherence}(a,b)). To examine the role of inter-site coupling, we compare the measured relative complementary-phase trajectories with simulations with and without spatial interactions (Fig.~\ref{fig:Fig4}(n)). The relative phases remain bounded in the experiment and in the coupled simulation. In the uncoupled simulation they drift progressively. In simulations without coupling, the mean pairwise sum-phase coherence changes little, whereas the complementary-phase coherence decreases substantially (Supplementary Fig.~\ref{fig:Figsm_spatial_coherence}(d)). These results thus distinguish pump-induced sum-phase locking from interaction-mediated synchronization of the winding complementary phase.

We test quasiperiodicity at this operating point through low-order frequency relations and the recurrence of the reconstructed phase trajectory (Fig.~\ref{fig:Fig5}). The independently extracted frequencies identify the sum-resonant family $\omega_1+\omega_2\simeq\Omega$, and every tested low-order combination independent of the sum condition has a mismatch above the experimental resolution (Fig.~\ref{fig:Fig5}(a)). The measured spectrum is organized by the same two fundamentals and their combination lines (Fig.~\ref{fig:Fig5}(b)). In the half-pump rotating coordinates, the reconstructed trajectory winds along the phase torus as the sum phase remains locked and the complementary phase evolves (Fig.~\ref{fig:Fig5}(c)). At lags \(\tau=2mT\), where the rotating-frame and laboratory-frame phase increments coincide modulo \(2\pi\), the return distance of the reconstructed trajectory shows repeated near-returns without closure over the recorded lags (Fig.~\ref{fig:Fig5}(d)). 

In a single unit cell pumped above threshold at the sum resonance, the pump fixes the sum of the two branch phases and leaves their difference free, and in an uncoupled array this freedom is independent at each site. Coupling preserves the freedom of a common shift and gives differences between neighbors a positive stiffness about the uniform state. Where this state is stable and relative-phase fluctuations stay small, the medium therefore shares one free phase, and its quasiperiodic response becomes a collective state. In the experiment we find a coherent complementary phase across the recorded sites, and in simulations the relative complementary phases drift once the coupling is removed. The frequency of the single-tone clock selects the resonance channel through which the lattice orders. Above threshold and at sufficiently low noise, a degenerate resonance yields a period-doubled time crystal, and the sum resonance yields a time quasicrystal whose two frequencies add up to the pump frequency, are incommensurate with it, and differ by an amount set by the detunings and nonlinear shifts of the medium.

The most direct next test acts on a single site, where a phase kick or local detuning would expose the coupled amplitude and relative-phase response, and weakening one link would show how relaxation and spatial coherence depend on the coupling. Longer arrays open a second route, since in our circuit model they admit unstable modes with nonzero wavevector whose saturated states remain to be mapped. While sum locking persists, diffusion of the shared complementary phase broadens both response lines, which can be tested by following linewidths and spatial phase correlations as the array grows. Extending the DTQC state to optical frequencies remains challenging, since it requires modulation sustained over many cycles with controlled dissipation and saturation, and we leave it to future work.

\section*{Acknowledgements}
This work was supported by the National Research Foundation of Korea (NRF) through the government of Korea (NRF-2022R1A2C301335313) and the Samsung Science and Technology Foundation (SSTF-BA2402-02). J.C. acknowledges support from AFOSR (YIP No. FA9550-25-1-0147) and the Terman Faculty Fellowship at Stanford University.

\bibliography{Submission_refs}
\clearpage

\clearpage 

\begin{widetext}

\setcounter{equation}{0}
\setcounter{figure}{0}
\setcounter{table}{0}

\renewcommand{\theequation}{S\arabic{equation}}
\renewcommand{\thefigure}{S\arabic{figure}}
\renewcommand{\thetable}{S\arabic{table}}

\renewcommand{\theHequation}{S\arabic{equation}}
\renewcommand{\theHfigure}{S\arabic{figure}}
\renewcommand{\theHtable}{S\arabic{table}}

{\centering
\large \textbf{Supplementary Information: Self-organized photonic time quasicrystal\\ from a single imposed clock} \\[2ex]  

\normalsize Minwook Kyung$^{1,*}$, Kyungmin Lee$^{1,*}$, Yung Kim$^{1}$, Eun-Gook Moon$^{1}$, Joonhee Choi$^{2}$ and Bumki Min$^{1,\dagger}$ \\[1ex] 
\fontsize{9pt}{10pt}\selectfont 
$^{1}$\textit{Department of Physics, Korea Advanced Institute of Science and Technology, Daejeon 34141, Republic of Korea} \\
$^{2}$\textit{Department of Electrical Engineering, Stanford University, Stanford, CA 94305, USA}\\
\textit{$^{\ast}$These authors contributed equally to this work.}\\
\textit{$^{\dagger}$ bmin@kaist.ac.kr}\\
}

\vspace{2ex}
\begin{itemize}

\item \textbf{Section S1. Maxwell--Lorentz equations of motion and hybrid carrier basis}
\begin{itemize}
    \item Maxwell--polarization equations
    \item First-order state-space form
    \item Unmodulated hybrid carriers
    \item Resonance conditions and selected carrier sectors
\end{itemize}

\item \textbf{Section S2. Circuit realization and projected slow Hamiltonian}
\begin{itemize}
    \item Flux--charge circuit model
    \item Reduced capacitive Hamiltonian
    \item Effective modulation and reduced circuit Hamiltonian
    \item Two-sector slow variables
    \item Hybrid basis and unified slow Hamiltonian
\end{itemize}

\item \textbf{Section S3. Branch decomposition: degenerate DTC channels and nondegenerate DTQC channel}
\begin{itemize}
    \item Linear Floquet branch landscape
    \item Branch I-A: DTC$_a$ from $\Omega\simeq 2\omega_a$
    \item Branch I-B: DTC$_b$ from $\Omega\simeq 2\omega_b$
    \item Branch II: DTQC from $\Omega\simeq\omega_a+\omega_b$
    \item Frozen-action phase reduction and deterministic sum-locking tongue
    \item Slow-flow equations and noise projection
    \item Validation of the rotating-wave reduction
\end{itemize}

\item \textbf{Section S4. Experimental implementation}
\begin{itemize}
    \item Experimental platform, measurement setup, and noise injection
    \item Signal processing and reconstruction of experimental DTQC observables
    \item Quantitative quasiperiodicity tests for the experimental DTQC state
    \item Spatial coherence and collective synchronization of the DTQC state
\end{itemize}

\end{itemize}
\clearpage

\renewcommand{\thesection}{S\arabic{section}}
\renewcommand{\thesubsection}{\Alph{subsection}}

\counterwithin{figure}{section}

\section{Maxwell--Lorentz equations of motion and hybrid carrier basis}
\label{sec:SI_ML_dynamical_model}

\subsection{Maxwell--polarization equations}
\label{subsec:SI_ML_EOM}

We consider a one-dimensional Maxwell--Lorentz medium with a nonlinear polarization coordinate \(P(z,t)\). We introduce a guided-wave coordinate \(\Phi(z,t)\) such that the local electric field is the generalized velocity
\begin{equation}
E(z,t)\equiv \dot{\Phi}(z,t).
\label{eq:SI_ML_E_def_compact}
\end{equation}
The displacement field is
\begin{equation}
D(z,t)=\varepsilon_b E(z,t)+P(z,t),
\label{eq:SI_ML_Dclosure_compact}
\end{equation}
and the corresponding displacement-like canonical momentum density is
\begin{equation}
Q(z,t)\equiv D(z,t)=\varepsilon_b\dot{\Phi}(z,t)+P(z,t),
\qquad
E(z,t)=\frac{Q(z,t)-P(z,t)}{\varepsilon_b}.
\label{eq:SI_ML_Q_def_compact}
\end{equation}

For a source-free one-dimensional guided mode with transverse fields, Maxwell's equations reduce to
\begin{equation}
\partial_z E(z,t)=-\mu\,\partial_t H(z,t),
\qquad
\partial_z H(z,t)=-\partial_t D(z,t)=-\partial_t Q(z,t).
\label{eq:SI_ML_Maxwell_1D_compact}
\end{equation}
Using \(E=\dot\Phi\) in the first equation gives
\(\partial_t(\partial_z\Phi+\mu H)=0\). For the oscillatory fields considered here, we set the time-independent integration constant to zero and obtain
\begin{equation}
H(z,t)=-\frac{1}{\mu}\partial_z\Phi(z,t).
\label{eq:SI_ML_H_from_Phi_compact}
\end{equation}
Substitution into the second Maxwell equation gives
\begin{equation}
\dot Q(z,t)=\frac{1}{\mu}\partial_z^2\Phi(z,t).
\label{eq:SI_ML_Qdot_compact}
\end{equation}
Equations~\eqref{eq:SI_ML_Q_def_compact} and \eqref{eq:SI_ML_Qdot_compact} provide the first-order guided-wave sector, with the physical fields recovered through
\begin{equation}
E=\frac{Q-P}{\varepsilon_b},
\qquad
H=-\frac{1}{\mu}\partial_z\Phi.
\label{eq:SI_ML_field_readout_compact}
\end{equation}

The polarization is modeled as a nonlinear Lorentz oscillator driven by the local electric field,
\begin{equation}
\ddot P+\gamma\dot P+\Omega_0^2(t)P+\beta P^3
-v_P^2\partial_z^2P
=
\chi E+\frac{1}{m}F(z,t).
\label{eq:SI_ML_P_EOM_compact}
\end{equation}
Here \(\gamma\) is the polarization damping rate, \(\beta\) is the Duffing nonlinearity, \(v_P\) is the polarization-coupling velocity, \(\chi\) is the field--polarization coupling, and \(F\) is an effective Langevin force. The parameter \(m\) fixes the canonical normalization of the polarization momentum and the corresponding normalization of the Langevin force term. Since \(P\) is normalized as the polarization entering \(D=\varepsilon_bE+P\), we use the convention \(\Pi\equiv m\dot P\) and \(m\chi=1\). This convention fixes \(m\) and \(\Pi\) for the chosen polarization coordinate and adds no physical constraint.

The imposed periodic drive modulates the polarization stiffness,
\begin{equation}
\Omega_0^2(t)=\omega_0^2\left[1+\delta\cos(\Omega t)\right],
\qquad
T=\frac{2\pi}{\Omega}.
\label{eq:SI_ML_stiffness_mod_compact}
\end{equation}
Thus the explicit time dependence enters through an internal material coordinate, and no instantaneous \(\varepsilon(t)\) is prescribed.

For the discrete dipole lattice used to connect with the experimental array, \(z_n=n d\) and \(P_n(t)\equiv P(z_n,t)\). With the standard central-difference convention,
\[
-v_P^2\partial_z^2P(z_n,t)
\ \longrightarrow
g\left(2P_n-P_{n+1}-P_{n-1}\right),
\qquad
g=\frac{v_P^2}{d^2}.
\]
Accordingly, Eq.~\eqref{eq:SI_ML_P_EOM_compact} becomes
\begin{equation}
\ddot P_n+\gamma\dot P_n+\Omega_0^2(t)P_n+\beta P_n^3
+g\left(2P_n-P_{n+1}-P_{n-1}\right)
=
\chi E_n+\frac{1}{m}F_n(t),
\label{eq:SI_ML_discrete_P_EOM_compact}
\end{equation}
where \(E_n(t)\) is the guided electromagnetic field evaluated at the \(n\)-th unit cell, itself a dynamical variable. The discrete equation is therefore a lattice realization of the same Maxwell--polarization medium rather than a separate lumped-oscillator model.

For the theoretical Maxwell--polarization model, we introduce a lattice-level Langevin force acting on the polarization sector. Its dimensional mean and covariance are defined by
\begin{equation}
\langle F_n(t)\rangle=0,
\qquad
\langle F_n(t)F_m(t')\rangle
=
2m\gamma D_{\rm phys}\,\delta_{nm}\delta(t-t') .
\label{eq:SI_ML_noise_discrete_dimensional}
\end{equation}
Here \(D_{\rm phys}\) is an effective dimensional noise scale conjugate to the polarization coordinate. It differs from the measured microwave noise spectral density \(S_{\mathrm{noise}}\), which is the externally injected broadband noise before projection through the waveguide, coupling network, and resonator response.

To nondimensionalize this polarization-force convention, we introduce \(\tau=\omega_*t\) and \(p_n=P_n/P_*\). We choose \(\omega_*=\omega_0\) and \(P_*=\omega_0/\sqrt{|\beta|}\), so that the dimensionless Duffing coefficient reduces to \(\operatorname{sgn}(\beta)\). Defining \(f_n(\tau)=F_n(t)/(mP_*\omega_*^2)\), Eq.~\eqref{eq:SI_ML_noise_discrete_dimensional} becomes
\begin{equation}
\langle f_n(\tau)f_m(\tau')\rangle
=
2\bar{\gamma}\mathcal D_{\mathrm{pol}}\,\delta_{nm}\delta(\tau-\tau'),
\qquad
\bar{\gamma}=\frac{\gamma}{\omega_*}.
\label{eq:SI_ML_noise_discrete_dimensionless}
\end{equation}
with
\begin{equation}
\mathcal D_{\mathrm{pol}}=\frac{D_{\mathrm{phys}}}{D_0},
\qquad
D_0=mP_*^2\omega_*^2
=
\frac{m\omega_0^4}{|\beta|}.
\label{eq:SI_ML_noise_D0}
\end{equation}
This normalization defines the noise intensity of the theoretical polarization-force convention.

The numerical phase-memory maps and switching statistics are obtained from the rotating-wave reduction of the effective circuit equations, which retains all normal modes of both bands and the pump-resonant terms, driven by independent additive charge noise, \(dQ_{s,n}=F_{Q_s,n}\,dt+\sqrt{2D_Q}\,dW_{s,n}\), projected onto the mode envelopes, where \(s\in\{0,r\}\), \(F_{Q_s,n}\) contains the deterministic Hamiltonian drift and linear damping, and the Wiener processes are independent between sites and charge sectors. All numerical noise levels are reported directly as the charge-diffusion coefficient \(D_Q\), with units of \(\mathrm{C^2/s}\). The simulated resonator voltages used for the phase-memory maps, line cuts, and Fig.~3(d) include additive white readout noise calibrated to the measured spectral noise floor; this readout noise does not enter the dynamics.

The polarization-force noise scale, the numerical charge-diffusion coefficient $D_Q$, and the experimentally injected spectral density $S_{\mathrm{noise}}$ are thus three distinct quantities; the propagation of the injected noise through the waveguide and resonator network is kept separate from $D_Q$.

\subsection{First-order state-space form}
\label{subsec:SI_ML_state_space_compact}

For the reference hybrid-carrier analysis, we use the homogeneous continuum Maxwell--Lorentz model in $k$-space. Floquet branch selection for the circuit realization is evaluated separately using the discrete circuit model in Sec.~\ref{subsec:SI_branch_linear_landscape}. With
\begin{equation}
\Phi(z,t)=\Phi_k(t)e^{ikz},
\qquad
Q(z,t)=Q_k(t)e^{ikz},
\qquad
P(z,t)=P_k(t)e^{ikz},
\label{eq:SI_ML_plane_wave_compact}
\end{equation}
the polarization stiffness becomes
\[
\Omega_k^2(t)\equiv\Omega_0^2(t)+v_P^2k^2 .
\]
Defining \(\Pi\equiv m\dot P\) and using the canonical normalization convention \(m\chi=1\), we collect the variables as
\begin{equation}
\mathbf a(k,t)=
\begin{bmatrix}
\Phi\\ Q\\ P\\ \Pi
\end{bmatrix}.
\label{eq:SI_ML_state_vector_compact}
\end{equation}
The linearized dynamics is
\begin{equation}
\frac{d}{dt}\mathbf a(k,t)=\mathsf M(k,t)\mathbf a(k,t),
\label{eq:SI_ML_first_order_compact}
\end{equation}
with
\begin{equation}
\mathsf M(k,t)=
\begin{bmatrix}
0 & \dfrac{1}{\varepsilon_b} & -\dfrac{1}{\varepsilon_b} & 0\\[6pt]
-\dfrac{k^2}{\mu} & 0 & 0 & 0\\[4pt]
0 & 0 & 0 & \dfrac{1}{m}\\[6pt]
0 & \dfrac{1}{\varepsilon_b} &
-m\Omega_k^2(t)-\dfrac{1}{\varepsilon_b}
& -\gamma
\end{bmatrix}.
\label{eq:SI_ML_M_matrix_compact}
\end{equation}
The physical fields are recovered through
\begin{equation}
E_k=\frac{Q_k-P_k}{\varepsilon_b},
\qquad
H_k=-\frac{ik}{\mu}\Phi_k .
\label{eq:SI_ML_readout_compact}
\end{equation}
The Duffing nonlinearity is local in real space and appears in \(k\)-space as the convolution \([P^3]_k\). We therefore use Eq.~\eqref{eq:SI_ML_M_matrix_compact} for the linear Maxwell--Lorentz reference analysis; the circuit-specific Floquet spectrum and nonlinear stochastic dynamics are treated separately in Supplementary Sections~S2 and S3.

\subsection{Unmodulated hybrid carriers}
\label{subsec:SI_ML_hybrid_carriers_compact}

The reference hybrid-carrier bands are obtained from the unmodulated linear Maxwell--polarization problem. We set \(\delta=0\), \(\beta=0\), \(F=0\), and temporarily neglect damping. Harmonic solutions then satisfy
\begin{equation}
\left[\omega^2-\omega_{\rm wg}^2(k)\right]
\left[\omega^2-\omega_{\rm res}^2(k)\right]
=
\omega_p^2\omega^2,
\label{eq:SI_ML_polariton_disp_compact}
\end{equation}
with
\begin{equation}
\omega_{\rm wg}^2(k)=\frac{k^2}{\mu\varepsilon_b},
\qquad
\omega_{\rm res}^2(k)=\omega_0^2+v_P^2k^2,
\qquad
\omega_p^2=\frac{\chi}{\varepsilon_b}=\frac{1}{m\varepsilon_b}.
\label{eq:SI_ML_bare_defs_compact}
\end{equation}
The last equality follows from the canonical normalization convention \(m\chi=1\). Equation~\eqref{eq:SI_ML_polariton_disp_compact} gives two positive-frequency hybrid branches,
\begin{equation}
\omega_{\mathrm L,\mathrm U}^2(k)
=
\frac{1}{2}
\left[
\omega_{\rm wg}^2(k)+\omega_{\rm res}^2(k)+\omega_p^2
\mp
\sqrt{
\left[\omega_{\rm wg}^2(k)+\omega_{\rm res}^2(k)+\omega_p^2\right]^2
-4\omega_{\rm wg}^2(k)\omega_{\rm res}^2(k)}
\right],
\label{eq:SI_ML_hybrid_branches_compact}
\end{equation}
where the minus sign gives the lower band \(\omega_{\mathrm L}(k)\) and the plus sign gives the upper band \(\omega_{\mathrm U}(k)\). 
In the weak-hybridization limit, the lower and upper bands reduce to the lower and upper of the two uncoupled dispersions,
\begin{equation}
    \omega_L^2(k)\to
\min\{\omega_{\rm wg}^2(k),\omega_{\rm res}^2(k)\},\qquad
\omega_U^2(k)\to
\max\{\omega_{\rm wg}^2(k),\omega_{\rm res}^2(k)\}.
\label{eq:SI_ML_weak_hybrid_limit_compact}
\end{equation}

On a discrete lattice, the corresponding bare dispersions are replaced by
\begin{equation}
\omega_{\rm wg}^2(k)
=
\frac{4}{\mu\varepsilon_b d^2}
\sin^2\left(\frac{k d}{2}\right),
\qquad
\omega_{\rm res}^2(k)
=
\omega_0^2+4g\sin^2\left(\frac{k d}{2}\right),
\label{eq:SI_ML_lattice_disp_compact}
\end{equation}
with \(g=v_P^2/d^2\) in the central-difference convention used in Eq.~\eqref{eq:SI_ML_discrete_P_EOM_compact}. The hybridization structure of Eq.~\eqref{eq:SI_ML_polariton_disp_compact} is unchanged after replacing the bare dispersions by Eq.~\eqref{eq:SI_ML_lattice_disp_compact}.

\begin{figure*}[htb]
  \centering
  \includegraphics[width=0.8\textwidth]{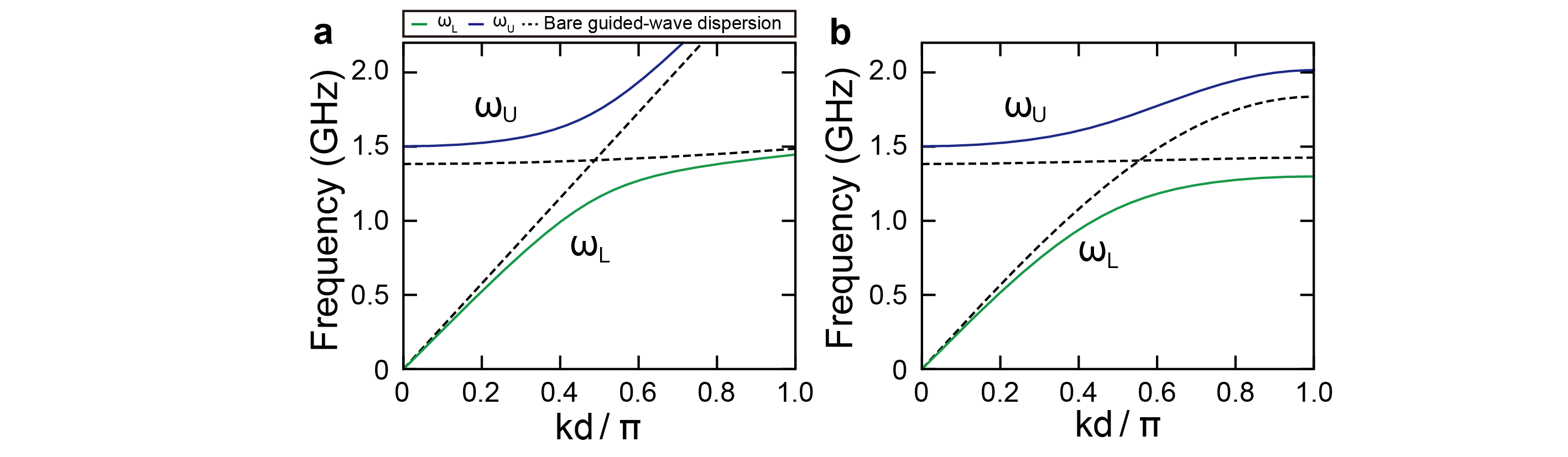}
  \caption{\textbf{Unmodulated hybrid dispersion.} (a,b) Positive-frequency hybrid branches \(\omega_{\mathrm L}(k)\) and \(\omega_{\mathrm U}(k)\) obtained from the unmodulated Maxwell--Lorentz model in the continuum and discrete-lattice descriptions. Dashed curves show the uncoupled guided-wave and polarization dispersions.}
  \label{fig:SI_hybrid_dispersion_participation}
\end{figure*}

\subsection{Resonance conditions and selected carrier sectors}
\label{subsec:SI_ML_participation_compact}

The driven normal form uses the unmodulated hybrid bands as carrier branches. We denote the lower and upper positive-frequency branches obtained in Eq.~\eqref{eq:SI_ML_hybrid_branches_compact} by
\[
\omega_{\mathrm L}(k),\qquad \omega_{\mathrm U}(k).
\]
These bands are hybrid Maxwell--polarization modes: their electromagnetic and polarization contents vary with \(k\), but the resonance conditions used below depend only on the carrier frequencies.

Here \(k_0\) denotes the resonant Bloch wave number, or the center of the narrow momentum window selected by the corresponding Floquet instability in a finite lattice.
The two resonance classes central to the driven problem are first expressed at the level of these unmodulated bands. A degenerate \(2{:}1\) channel selects one band when
\begin{equation}
\Omega\simeq 2\omega_\nu(k_0),
\qquad
\nu\in\{\mathrm L,\mathrm U\}.
\label{eq:SI_ML_degenerate_condition_compact}
\end{equation}
This condition gives the single-branch DTC channels. The nondegenerate sum-resonant channel selects one carrier from each band when
\begin{equation}
\Omega\simeq \omega_{\mathrm L}(k_0)+\omega_{\mathrm U}(k_0).
\label{eq:SI_ML_sum_condition_compact}
\end{equation}
For a spatially uniform temporal modulation, momentum is conserved; in a real-field representation, Eq.~\eqref{eq:SI_ML_sum_condition_compact} corresponds to pairing opposite-momentum components \(\pm k_0\) with the same band frequencies.

In the branch-reduced slow Hamiltonian used in the main text, the symbols \(a\) and \(b\) denote the selected near-resonant carrier sectors. Thus, for the DTQC operating point,
\begin{equation}
\omega_a\simeq\omega_{\mathrm L}(k_0),
\qquad
\omega_b\simeq\omega_{\mathrm U}(k_0),
\label{eq:SI_band_to_selected_carriers}
\end{equation}
whereas \(\omega_{\mathrm L,\mathrm U}(k)\) denote the underlying unmodulated hybrid bands. The effective coefficients of the slow Hamiltonian are obtained by projecting the driven lattice dynamics onto these carrier sectors and retaining the near-resonant terms.

\section{Circuit realization and projected slow Hamiltonian}
\label{sec:SI_circuit_slow_hamiltonian}

The experimental lattice is represented by a flux--charge circuit model. This formulation gives the element-level dictionary between the capacitances, inductances, varactor modulation, nonlinear charge response, and the projected slow-Hamiltonian coefficients used in the branch decomposition. We start from a bus node capacitively coupled to a nonlinear resonator node and project the time-dependent quadratic channels onto the hybrid basis obtained by diagonalizing the unmodulated circuit Hamiltonian.

\subsection{Flux--charge circuit model}
\label{subsec:SI_circuit_flux_charge_model}

Each unit cell contains a bus node with flux $\Phi_n(t)$ and a resonator node with flux $\Psi_n(t)$. The corresponding node voltages are
\begin{equation}
V_{0,n}=\dot\Phi_n,
\qquad
V_{r,n}=\dot\Psi_n.
\label{eq:SI_circuit_node_voltages}
\end{equation}
The bus node is shunted by a capacitance $C_0$, the bus and resonator nodes are coupled by a capacitance $C_g$, and the resonator node contains a time-modulated nonlinear capacitance. The resonator charge energy is written as
\begin{equation}
U_r(q_{r,n};t)
=
\frac{q_{r,n}^2}{2C_{r,L}(t)}
+
\frac{q_{r,n}^4}{4C_{r,{\rm NL}}},
\qquad
C_{r,L}(t)
=
\frac{C_{r,0}}{1+\delta_r\cos(\Omega t)}.
\label{eq:SI_circuit_Ur}
\end{equation}
The inductive network is separated into a guided bus sector and a resonator sector. 
The bus sector consists of the on-site shunt inductors \(L_s\) and nearest-neighbor series inductors \(L_0\), giving the discrete-gradient energy in Eq.~\eqref{eq:SI_circuit_Lagrangian}. 
The resonator sector consists of the resonator self-inductors \(L_r\) and an effective nearest-neighbor inductive coupling parameter \(M\). 
We parameterize the resonator inductive coupling through the effective nearest-neighbor inverse-inductance operator
\begin{equation}
\mathbf L_r^{-1}
=
\frac{1}{L_r}\mathbf I-k_r\mathbf L_2,
\qquad
k_r\equiv\frac{M}{L_r^2},
\label{eq:SI_circuit_Lrinv}
\end{equation}
where $\mathbf L_2$ is the second-difference discrete Laplacian and $M$ is an effective coupling parameter fitted to the hybrid-band dispersion. This operator has the form of the first-order expansion of $(L_r\mathbf I+M\mathbf L_2)^{-1}$.

We use a first-order flux--charge representation. 
Here \(\Phi_n\) and \(\Psi_n\) are, respectively, the bus flux and resonator flux in the \(n\)-th unit cell. 
The branch charges \(q_{0,n}\), \(q_{g,n}\), and \(q_{r,n}\) denote the charges on the bus capacitance \(C_0\), the bus--resonator coupling capacitor \(C_g\), and the resonator varactor branch, respectively. 
The orientation of \(q_{g,n}\) is chosen from the bus node to the resonator node, so that its first-order term is \(q_{g,n}(\dot\Phi_n-\dot\Psi_n)\). 
The Lagrangian is then
\begin{align}
\mathcal L
&=
\sum_n
\left[
q_{0,n}\dot\Phi_n-\frac{q_{0,n}^2}{2C_0}
+q_{g,n}(\dot\Phi_n-\dot\Psi_n)-\frac{q_{g,n}^2}{2C_g}
+q_{r,n}\dot\Psi_n-U_r(q_{r,n};t)
\right]
\nonumber\\
&\quad
-\sum_n
\left[
\frac{\Phi_n^2}{2L_s}
+
\frac{(\Phi_{n+1}-\Phi_n)^2}{2L_0}
\right]
-\frac{1}{2}\Psi^{\mathsf T}\mathbf L_r^{-1}\Psi .
\label{eq:SI_circuit_Lagrangian}
\end{align}
The canonical charges conjugate to \(\Phi_n\) and \(\Psi_n\) are therefore
\begin{equation}
Q_{0,n}=q_{0,n}+q_{g,n},
\qquad
Q_{r,n}=q_{r,n}-q_{g,n}.
\label{eq:SI_circuit_canonical_charges}
\end{equation}

The independent circuit parameters entering Eqs.~\eqref{eq:SI_circuit_Ur}, \eqref{eq:SI_circuit_Lrinv}, and \eqref{eq:SI_circuit_Lagrangian} are summarized in Table~\ref{tab:circuit_bare_params}. The primed capacitances and effective modulation depths used below are derived from these bare circuit quantities.


\begin{table*}[htb]
\centering
\caption{\textbf{Effective circuit parameters and modulation settings.} The listed values specify the effective numerical model in Sec.~\ref{sec:SI_circuit_slow_hamiltonian}. Fixed circuit parameters are shared by Figs.~2 and 3. The drive frequencies are \(\Omega/(2\pi)=2.98\) GHz for DTC and \(2.47\) GHz for DTQC. The effective modulation depth is \(\delta_{\mathrm{eff}}=\alpha_0\delta_r\), with \(\alpha_0\simeq0.63\) determined by the circuit capacitances.}
\label{tab:circuit_bare_params}
\footnotesize
\renewcommand{\arraystretch}{1.18}
\begin{tabular}{@{}c p{0.48\textwidth} c c@{}}
\toprule
\textbf{Symbol} & \multicolumn{1}{c}{\textbf{Meaning}} & \textbf{Value} & \textbf{Unit} \\
\midrule
\(C_0\) & Bus-node shunt capacitance to ground & \(2.140\times10^{-10}\) & \(\mathrm{F}\) \\[0.8ex]
\(C_g\) & Bus--resonator coupling capacitance & \(7.634\times10^{-11}\) & \(\mathrm{F}\) \\[0.8ex]
\(C_{r,0}\) & Unmodulated linear resonator capacitance & \(9.775\times10^{-11}\) & \(\mathrm{F}\) \\[0.8ex]
\(C_{r,\mathrm{NL}}\) & Nonlinear-capacitance scale of the quartic resonator charge energy & \(1.225\times10^{-30}\) & \(\mathrm{F}\,\mathrm{C}^{2}\) \\[0.8ex]
\(\Omega\) & Modulation angular frequency & Branch-selected & \(\mathrm{rad/s}\) \\[0.8ex]
\(\delta_r\) & Bare modulation depth in \(C_{r,L}(t)=C_{r,0}/[1+\delta_r\cos(\Omega t)]\) & Swept or fixed & \(1\) \\[0.8ex]
\(L_s\) & On-site bus shunt inductance & \(8.239\times10^{-11}\) & \(\mathrm{H}\) \\[0.8ex]
\(L_0\) & Nearest-neighbour bus series inductance & \(4.116\times10^{-10}\) & \(\mathrm{H}\) \\[0.8ex]
\(L_r\) & Resonator self-inductance & \(8.239\times10^{-11}\) & \(\mathrm{H}\) \\[0.8ex]
\(M\) & Effective nearest-neighbour inductive coupling parameter & \(1.649\times10^{-11}\) & \(\mathrm{H}\) \\[0.8ex]
\(R_0=R_r\) & Assumed damping resistances in the numerical integrations & \(0.01\) & \(\Omega\) \\
\bottomrule
\end{tabular}
\renewcommand{\arraystretch}{1.0}
\end{table*}

\subsection{Reduced capacitive Hamiltonian}
\label{subsec:SI_circuit_reduced_capacitive_H}

Eliminating the internal capacitor charges gives the quadratic capacitive Hamiltonian
\begin{equation}
H_C(t)
=
\frac{1}{2}
\sum_n
\begin{bmatrix}
Q_{0,n} & Q_{r,n}
\end{bmatrix}
\mathsf C_{\rm eff}^{-1}(t)
\begin{bmatrix}
Q_{0,n}\\ Q_{r,n}
\end{bmatrix}.
\label{eq:SI_circuit_HC_matrix}
\end{equation}
At zero modulation, define
\begin{equation}
\Delta_0
=
C_0C_{r,0}+C_g(C_0+C_{r,0}).
\label{eq:SI_circuit_Delta0}
\end{equation}
The static effective capacitances are
\begin{equation}
\frac{1}{C_0'}
=
\frac{C_g+C_{r,0}}{\Delta_0},
\qquad
\frac{1}{C_r'}
=
\frac{C_0+C_g}{\Delta_0},
\qquad
\frac{1}{C_g'}
=
\frac{C_g}{\Delta_0}.
\label{eq:SI_circuit_Ceff_defs}
\end{equation}
Thus
\begin{equation}
H_C^{(0)}
=
\sum_n
\left[
\frac{Q_{0,n}^2}{2C_0'}
+
\frac{Q_{r,n}^2}{2C_r'}
+
\frac{Q_{0,n}Q_{r,n}}{C_g'}
\right].
\label{eq:SI_circuit_HC_static}
\end{equation}

The nonlinear capacitor gives
\begin{equation}
H_{\rm NL}
=
\sum_n
\frac{q_{r,n}^4}{4C_{r,{\rm NL}}}.
\label{eq:SI_circuit_HNL_exact}
\end{equation}
To leading order in the modulation depth and in the varactor nonlinearity, the unmodulated linear capacitance network gives
\begin{equation}
q_{r,n}
\simeq
\alpha_0 Q_{r,n}+\zeta_0 Q_{0,n},
\qquad
\alpha_0=\frac{C_{r,0}(C_0+C_g)}{\Delta_0},
\qquad
\zeta_0=\frac{C_gC_{r,0}}{\Delta_0},
\label{eq:SI_circuit_qr_weakload}
\end{equation}
so that
\begin{equation}
H_{\rm NL}
\simeq
\sum_n
\frac{
\left(\alpha_0 Q_{r,n}+\zeta_0 Q_{0,n}\right)^4
}{4C_{r,{\rm NL}}}.
\label{eq:SI_circuit_HNL_weakload}
\end{equation}

\subsection{Effective modulation and reduced circuit Hamiltonian}
\label{subsec:SI_circuit_weak_modulation}

Expanding the inverse-capacitance matrix to first order in $\delta_r$ gives
\begin{align}
\frac{1}{C_r'(t)}
&\simeq
\frac{1}{C_r'}
\left[1+\delta_r'\cos(\Omega t)\right],
\nonumber\\
\frac{1}{C_g'(t)}
&\simeq
\frac{1}{C_g'}
\left[1+\delta_r'\cos(\Omega t)\right],
\nonumber\\
\frac{1}{C_0'(t)}
&\simeq
\frac{1}{C_0'}
\left[1+\delta_0'\cos(\Omega t)\right],
\label{eq:SI_circuit_effective_mod_depths}
\end{align}
where
\begin{equation}
\delta_r'
=
\delta_r
\frac{C_{r,0}(C_0+C_g)}{\Delta_0},
\qquad
\delta_0'
=
\delta_r
\frac{C_{r,0}C_g^2}{\Delta_0(C_g+C_{r,0})}.
\label{eq:SI_circuit_delta_prime_defs}
\end{equation}

The symbol \(\delta_r\) denotes the bare resonator-capacitance modulation depth applied in the circuit. The quantities \(\delta_r'\) and \(\delta_0'\) are the corresponding effective inverse-capacitance modulation depths after eliminating the internal capacitor charges. Because the resonator stiffness is proportional to the inverse capacitance, the Maxwell--Lorentz stiffness-modulation depth \(\delta\) used in the main text should be interpreted as an effective modulation depth after this circuit reduction. When the resonator-sector stiffness modulation is used as the reference, we identify
\begin{equation}
\delta \equiv \delta_{\rm eff}=\delta_r' .
\label{eq:SI_delta_eff_identification}
\end{equation}

The reduced circuit Hamiltonian becomes
\begin{align}
H_{\rm circ}(t)
&=
\sum_n
\Bigg[
\frac{Q_{0,n}^2}{2C_0'}
\left(1+\delta_0'\cos\Omega t\right)
+
\left(
\frac{Q_{r,n}^2}{2C_r'}
+
\frac{Q_{0,n}Q_{r,n}}{C_g'}
\right)
\left(1+\delta_r'\cos\Omega t\right)
\nonumber\\
&\qquad
+
\frac{\Phi_n^2}{2L_s}
+
\frac{(\Phi_{n+1}-\Phi_n)^2}{2L_0}
\Bigg]
+
\frac{1}{2}\Psi^{\mathsf T}\mathbf L_r^{-1}\Psi
+
\sum_n
\frac{
\left(\alpha_0 Q_{r,n}+\zeta_0 Q_{0,n}\right)^4
}{4C_{r,{\rm NL}}}.
\label{eq:SI_circuit_H_full_reduced}
\end{align}
Equation~\eqref{eq:SI_circuit_H_full_reduced} defines the effective circuit model used throughout the following analysis, with $\delta_{\rm eff}$ specifying the resonator-sector inverse-capacitance modulation depth. The relation $\delta_{\rm eff}=\alpha_0\delta_r$ provides its leading-order correspondence to the bare varactor modulation. The bus coordinate $\Phi_n$ represents the guided-wave sector, while the resonator coordinate $\Psi_n$ provides the dynamical polarization sector.

\subsection{Two-sector slow variables}
\label{subsec:SI_circuit_AA_before_hybrid}

We introduce narrow-band action--angle variables in a frame rotating at $\Omega/2$. Let $\omega_{\rm wg}$ and $\omega_r$ denote the selected unmodulated bus-like and resonator-like carrier frequencies. The corresponding complex amplitudes are
\begin{equation}
c_{0,n}=\sqrt{J_{0,n}}e^{-i\tilde\theta_{0,n}},
\qquad
c_{1,n}=\sqrt{J_{1,n}}e^{-i\tilde\theta_{1,n}},
\qquad
\tilde\theta_{\ell,n}=\theta_{\ell,n}-\frac{\Omega t}{2}.
\label{eq:SI_circuit_c01_defs}
\end{equation}
The unmodulated number-conserving onsite quadratic sector is
\begin{equation}
\bar H_{\rm nc}^{(2)}
=
\sum_n
\begin{bmatrix}
c_{0,n}^\ast & c_{1,n}^\ast
\end{bmatrix}
\begin{bmatrix}
\Delta_{\rm wg} & -G/2\\
-G/2 & \Delta
\end{bmatrix}
\begin{bmatrix}
c_{0,n}\\ c_{1,n}
\end{bmatrix},
\label{eq:SI_circuit_H2_nc}
\end{equation}
with
\begin{equation}
\Delta_{\rm wg}=\omega_{\rm wg}-\frac{\Omega}{2},
\qquad
\Delta=\omega_r-\frac{\Omega}{2}.
\label{eq:SI_circuit_detunings_before_hybrid}
\end{equation}
The coupling scale $G$ is inherited from the static bus--resonator capacitive coupling.

The time-periodic modulation of the quadratic capacitive terms produces an effective bus-sector parametric coefficient $h_{\rm wg}$, a resonator-sector coefficient $h$, and a modulation-induced inter-sector coefficient $G_+$. The quadratic pump sector can be written as
\begin{align}
\bar H_{\rm pump}^{(2)}
&=
\sum_n
\left[
\frac{h_{\rm wg}}{2}J_{0,n}\cos(2\tilde\theta_{0,n})
+
\frac{h}{2}J_{1,n}\cos(2\tilde\theta_{1,n})
\right]
\nonumber\\
&\quad
-
G_+
\sum_n
\sqrt{J_{0,n}J_{1,n}}
\cos(\tilde\theta_{0,n}+\tilde\theta_{1,n}).
\label{eq:SI_circuit_Hpump_two_sector}
\end{align}
The onsite nonlinearity and spatial phase stiffnesses are summarized by
\begin{equation}
\bar H_{\rm Kerr}
=
\sum_n
\frac{\Lambda}{2}J_{1,n}^2,
\label{eq:SI_circuit_Kerr_before_hybrid}
\end{equation}
and
\begin{align}
\bar H_{\rm int}
&=
-\sum_n
K_{\rm wg}
\sqrt{J_{0,n}J_{0,n+1}}
\cos(\tilde\theta_{0,n}-\tilde\theta_{0,n+1})
\nonumber\\
&\quad
-\sum_n
K
\sqrt{J_{1,n}J_{1,n+1}}
\cos(\tilde\theta_{1,n}-\tilde\theta_{1,n+1}).
\label{eq:SI_circuit_stiffness_before_hybrid}
\end{align}

\subsection{Hybrid basis and unified slow Hamiltonian}
\label{subsec:SI_circuit_hybrid_slow_H}

The unmodulated quadratic sector in Eq.~\eqref{eq:SI_circuit_H2_nc} is diagonalized by the real rotation
\begin{equation}
\begin{bmatrix}
c_{0,n}\\ c_{1,n}
\end{bmatrix}
=
\begin{bmatrix}
\cos\psi & \sin\psi\\
-\sin\psi & \cos\psi
\end{bmatrix}
\begin{bmatrix}
c_{a,n}\\ c_{b,n}
\end{bmatrix},
\qquad
\tan(2\psi)=\frac{G}{\Delta_{\rm wg}-\Delta}.
\label{eq:SI_circuit_hybrid_rotation}
\end{equation}
In this projected slow description, \(a\) and \(b\) label the two carrier sectors drawn from the lower and upper hybrid bands. We define
\begin{equation}
c_{\alpha,n}=\sqrt{J_{\alpha,n}}e^{-i\tilde\theta_{\alpha,n}},
\qquad
\alpha\in\{a,b\}.
\label{eq:SI_circuit_hybrid_AA_defs}
\end{equation}
The hybrid detunings are
\begin{equation}
\Delta_{a,b}
=
\frac{\Delta_{\rm wg}+\Delta}{2}
\mp
\sqrt{
\left(\frac{\Delta_{\rm wg}-\Delta}{2}\right)^2
+
\left(\frac{G}{2}\right)^2
},
\label{eq:SI_circuit_Delta_ab}
\end{equation}
with hybrid carrier frequencies
\begin{equation}
\omega_\alpha=\Delta_\alpha+\frac{\Omega}{2}.
\label{eq:SI_circuit_omega_alpha}
\end{equation}

Projecting the quadratic pump sector onto the hybrid basis gives
\begin{equation}
\bar H_{\rm pump}^{(2)}
=
\sum_n
\left[
\sum_{\alpha=a,b}
\frac{h_\alpha}{2}
J_{\alpha,n}\cos(2\tilde\theta_{\alpha,n})
-
g_{ab}
\sqrt{J_{a,n}J_{b,n}}
\cos(\tilde\theta_{a,n}+\tilde\theta_{b,n})
\right],
\label{eq:SI_circuit_projected_pump}
\end{equation}
where
\begin{align}
h_a &= h_{\rm wg}\cos^2\psi + h\sin^2\psi + G_+\sin(2\psi),
\nonumber\\
h_b &= h_{\rm wg}\sin^2\psi + h\cos^2\psi - G_+\sin(2\psi),
\nonumber\\
g_{ab} &= \frac{h-h_{\rm wg}}{2}\sin(2\psi) + G_+\cos(2\psi).
\label{eq:SI_circuit_hab_gab}
\end{align}
The Kerr and stiffness coefficients become
\begin{equation}
\Lambda_{aa}=\Lambda\sin^4\psi,
\qquad
\Lambda_{bb}=\Lambda\cos^4\psi,
\qquad
\Lambda_{ab}=2\Lambda\sin^2\psi\cos^2\psi,
\label{eq:SI_circuit_eps_coeffs}
\end{equation}
and
\begin{equation}
K_a = K_{\rm wg}\cos^2\psi+K\sin^2\psi,
\qquad
K_b = K_{\rm wg}\sin^2\psi+K\cos^2\psi.
\label{eq:SI_circuit_Kab}
\end{equation}
Off-diagonal gradient couplings are higher order in the narrow-band projection and are neglected at this order.

Collecting the projected terms gives the slow Hamiltonian used in the branch decomposition
\begin{align}
\bar{\mathcal H} &=
\sum_{n,\alpha=a,b}
\left[ \Delta_\alpha J_{\alpha,n} + \frac{\Lambda_{\alpha\alpha}}{2} J_{\alpha,n}^2 \right]
+
\sum_n \Lambda_{ab}J_{a,n}J_{b,n}
\nonumber\\
&\quad
-
\sum_{n,\alpha=a,b} K_\alpha \sqrt{J_{\alpha,n}J_{\alpha,n+1}} \cos(\tilde\theta_{\alpha,n}-\tilde\theta_{\alpha,n+1})
\nonumber\\
&\quad
+
\sum_{n,\alpha=a,b} \frac{h_\alpha}{2} J_{\alpha,n} \cos(2\tilde\theta_{\alpha,n})
-
g_{ab}
\sum_n \sqrt{J_{a,n}J_{b,n}} \cos(\tilde\theta_{a,n}+\tilde\theta_{b,n}).
\label{eq:SI_circuit_H_unified}
\end{align}

The slow reduction assumes that the characteristic envelope rates are small compared with the drive frequency:
\begin{equation}
|\Delta_\alpha|,\ |K_\alpha|,\ |h_\alpha|,\ |g_{ab}|,\ 
|\Lambda_{\alpha\alpha}|\bar J_\alpha,\ 
|\Lambda_{ab}|\bar J_\beta
\ll\Omega,
\qquad \alpha\ne\beta.
\label{eq:SI_slow_reduction_condition}
\end{equation}
Nonsecular terms then average out over the fast oscillations.

\section{Branch decomposition: degenerate DTC channels and nondegenerate DTQC channel}
\label{sec:SI_branch_decomposition}

The unified slow Hamiltonian in Eq.~\eqref{eq:SI_circuit_H_unified} contains the quadratic pump channels relevant to the driven lattice: two intrabranch degenerate locking terms proportional to \(h_a\) and \(h_b\), and one interbranch sum-phase locking term proportional to \(g_{ab}\). These terms define the branch taxonomy used in the main text:
\begin{equation}
{\rm DTC}_a:\ \Omega\simeq 2\omega_a,
\qquad
{\rm DTC}_b:\ \Omega\simeq 2\omega_b,
\qquad
{\rm DTQC}:\ \Omega\simeq \omega_a+\omega_b.
\label{eq:SI_branch_conditions_summary}
\end{equation}

\subsection{Linear Floquet branch landscape}
\label{subsec:SI_branch_linear_landscape}

The same resonance conditions appear in the time-periodic linearized dynamics as parametric-instability channels. At the level of the underlying unmodulated hybrid bands, degenerate pumping produces instability intervals near
\begin{equation}
2\omega_\nu(k)\simeq \Omega,
\qquad
\nu\in\{\mathrm L,\mathrm U\},
\label{eq:SI_branch_linear_degenerate}
\end{equation}
whereas the nondegenerate channel produces instability tongues near
\begin{equation}
\omega_{\mathrm L}(k)+\omega_{\mathrm U}(k)\simeq \Omega.
\label{eq:SI_branch_linear_sum}
\end{equation}
The former corresponds to excitation of a single hybrid band near half the pump frequency, whereas the latter corresponds to simultaneous excitation of one carrier from each band, with their sum phase locked to the drive.

Figure~\ref{fig:SI_hybrid_dispersion_participation} presents the Maxwell--Lorentz reference bands, while Fig.~\ref{fig:SI_Floquet_branch_landscape} resolves the corresponding parametric-instability channels in the effective circuit model. The discrete bus includes the on-site shunt inductance \(L_s\), which sets a finite cutoff \(\omega_{\rm wg}(0)=1/\sqrt{L_sC_0'}\), and the nearest-neighbour series inductance \(L_0\), which sets its dispersion. The cutoff lifts \(\omega_{\rm L}(0)\) to a finite frequency and places the sum-resonant pair \(\omega_{\rm L}(k_0)+\omega_{\rm U}(k_0)\simeq\Omega\) near \(k_0\simeq0\). For a Floquet multiplier \(\rho(k)\) of the one-period evolution matrix, we define
\begin{equation}
\mu(k)=\frac{1}{T}\log\rho(k).
\label{eq:SI_Floquet_exponent_definition}
\end{equation}
Panels (b,c) show the quasifrequencies \(\operatorname{Im}\mu/\Omega\), folded modulo the drive frequency, and panels (d,e) show the normalized growth rates \(\operatorname{Re}\mu/\Omega\).

\begin{figure*}[htb]
  \centering
  \includegraphics[width=0.80\textwidth]{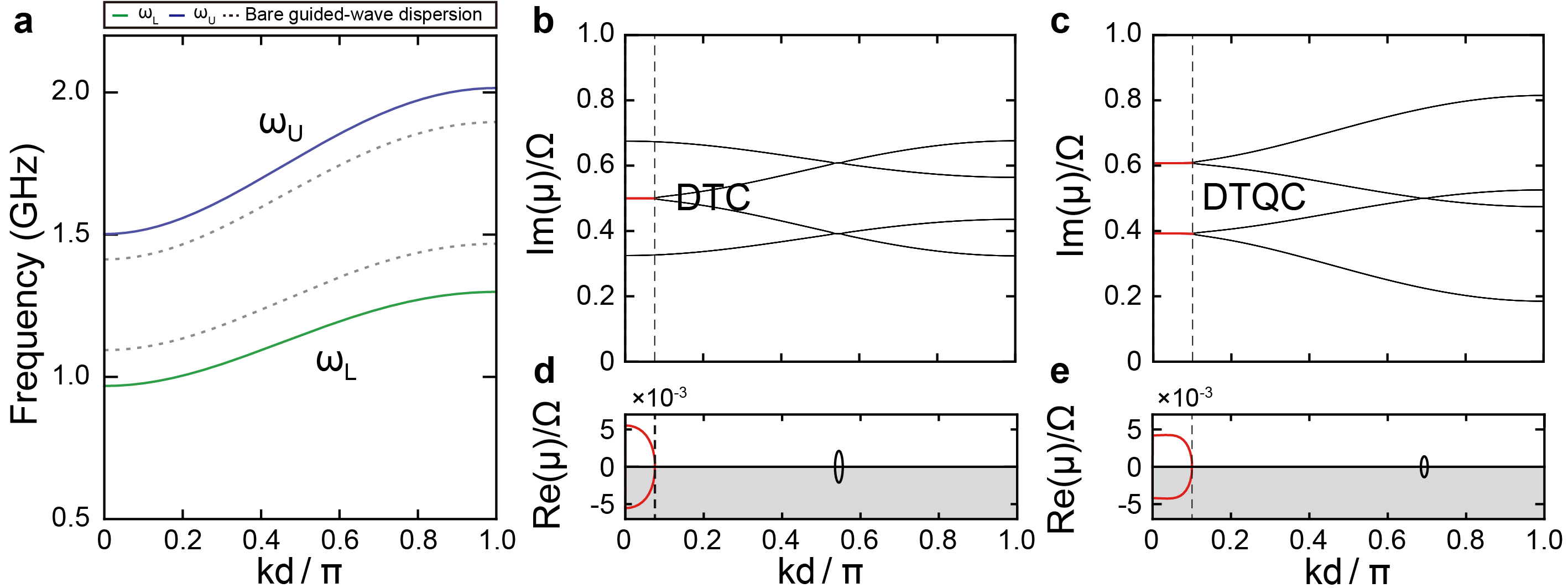}
    \caption{\textbf{Floquet branch selection in the circuit realization.}
    (a) Unmodulated lower and upper hybrid bands \(\omega_{\mathrm L}(k)\) and \(\omega_{\mathrm U}(k)\) of the discrete guided-wave--resonator circuit model. The dotted curves show the bare guided-wave (lower) and resonator (upper) dispersions without the cross term \(Q_0Q_r/C_g'\) of Eq.~\eqref{eq:SI_circuit_HC_static}; the on-site shunt \(L_s\) gives the guided-wave branch a finite cutoff at \(k=0\), and the lattice periodicity flattens both branches at the zone edge.
    (b,c) Floquet quasifrequencies \(\mathrm{Im}(\mu)/\Omega\) folded into the Floquet zone for the degenerate DTC and nondegenerate DTQC pumping conditions. Red segments and dashed vertical lines mark the primary near-resonant momentum windows.
    (d,e) Corresponding growth rates $\mathrm{Re}(\mu)/\Omega$ of the lossless linearized circuit. Positive growth identifies parametric-instability windows: the degenerate channels are selected near $2\omega_\nu(k)\simeq\Omega$, with $\nu\in\{\mathrm L,\mathrm U\}$, while the nondegenerate DTQC channel is selected near $\omega_{\mathrm L}(k)+\omega_{\mathrm U}(k)\simeq\Omega$. Weaker positive-growth features at finite momenta indicate secondary Floquet momentum gaps, distinct from the primary instability windows highlighted in red. The carrier frequencies associated with the primary windows are denoted by $\omega_a$ and $\omega_b$ in the branch-reduced slow Hamiltonian.}

  \label{fig:SI_Floquet_branch_landscape}
\end{figure*}

\subsection{Branch I-A: DTC$_a$ from $\Omega\simeq 2\omega_a$}
\label{subsec:SI_branch_DTCa}

The first degenerate channel is obtained by tuning the drive near the parametric resonance of hybrid branch $a$,
\begin{equation}
\Omega=2\omega_a+\Delta_{2,a},
\qquad
|\Delta_{2,a}|\ll \omega_a.
\label{eq:SI_DTCa_condition}
\end{equation}
Equivalently, the rotating-frame detuning
\begin{equation}
\Delta_a=\omega_a-\frac{\Omega}{2}
\label{eq:SI_DTCa_Deltaa}
\end{equation}
is small. If the complementary branch is off resonant,
\begin{equation}
|\Delta_b|
\gg
\left\{
|K_b|,\ |h_b|,\ |g_{ab}|
\right\},
\label{eq:SI_DTCa_offres_condition}
\end{equation}
branch $b$ can be eliminated to leading order. The effective Hamiltonian is
\begin{equation}
\bar{\mathcal H}_{{\rm DTC},a}
=
\sum_n
\left[
\Delta_a J_{a,n}
+
\frac{\Lambda_{aa}}{2}J_{a,n}^2
+
\frac{h_a}{2}J_{a,n}\cos(2\tilde\theta_{a,n})
\right]
-
\sum_n
K_a
\sqrt{J_{a,n}J_{a,n+1}}
\cos(\tilde\theta_{a,n}-\tilde\theta_{a,n+1}).
\label{eq:SI_H_DTCa}
\end{equation}
The term proportional to $h_a$ pins $2\tilde\theta_{a,n}$ and produces two symmetry-related subharmonic phase sectors. In the driven-dissipative dynamics, damping and nonlinear saturation convert this parametric instability into a period-doubled attractor.

\subsection{Branch I-B: DTC$_b$ from $\Omega\simeq 2\omega_b$}
\label{subsec:SI_branch_DTCb}

The second degenerate channel is obtained by tuning the drive near the parametric resonance of hybrid branch $b$,
\begin{equation}
\Omega=2\omega_b+\Delta_{2,b},
\qquad
|\Delta_{2,b}|\ll \omega_b.
\label{eq:SI_DTCb_condition}
\end{equation}
The rotating-frame detuning
\begin{equation}
\Delta_b=\omega_b-\frac{\Omega}{2}
\label{eq:SI_DTCb_Deltab}
\end{equation}
is then small. If branch $a$ is off resonant,
\begin{equation}
|\Delta_a|
\gg
\left\{
|K_a|,\ |h_a|,\ |g_{ab}|
\right\},
\label{eq:SI_DTCb_offres_condition}
\end{equation}
the slow Hamiltonian reduces to
\begin{equation}
\bar{\mathcal H}_{{\rm DTC},b}
=
\sum_n
\left[
\Delta_b J_{b,n}
+
\frac{\Lambda_{bb}}{2}J_{b,n}^2
+
\frac{h_b}{2}J_{b,n}\cos(2\tilde\theta_{b,n})
\right]
-
\sum_n
K_b
\sqrt{J_{b,n}J_{b,n+1}}
\cos(\tilde\theta_{b,n}-\tilde\theta_{b,n+1}).
\label{eq:SI_H_DTCb}
\end{equation}
In the weak-mixing limit, branch $b$ is predominantly resonator-like, with $\omega_b\rightarrow\omega_r$, $h_b\rightarrow h$, $K_b\rightarrow K$, and $\Lambda_{bb}\rightarrow\Lambda$. This is the branch most directly connected to the polarization-sector period-doubling mechanism.

\subsection{Branch II: DTQC from $\Omega\simeq\omega_a+\omega_b$}
\label{subsec:SI_branch_DTQC}

The DTQC branch is selected by the nondegenerate sum resonance
\begin{equation}
\Omega\simeq\omega_a+\omega_b.
\label{eq:SI_DTQC_condition}
\end{equation}
In the $\Omega/2$ rotating frame, this is equivalent to
\begin{equation}
\Delta_a+\Delta_b\simeq0.
\label{eq:SI_DTQC_Delta_sum}
\end{equation}
Near this condition, the dominant resonant pump channel is the interbranch term proportional to \(g_{ab}\). The intrabranch \(2{:}1\) terms proportional to \(h_a\) and \(h_b\) are off resonant in this nondegenerate sum-resonant branch and are therefore omitted from the branch-reduced DTQC Hamiltonian. If an additional low-order commensurability brought either intrabranch term into resonance, it would provide an extra phase-locking channel and could collapse the torus into a finite-period orbit. The effective two-branch Hamiltonian is
\begin{align}
\bar{\mathcal H}_{\rm DTQC}
&=
\sum_n
\left[
\Delta_a J_{a,n}
+
\Delta_b J_{b,n}
+
\frac{\Lambda_{aa}}{2}J_{a,n}^2
+
\frac{\Lambda_{bb}}{2}J_{b,n}^2
+
\Lambda_{ab}J_{a,n}J_{b,n}
\right]
\nonumber\\
&\quad
-
\sum_n
K_a
\sqrt{J_{a,n}J_{a,n+1}}
\cos(\tilde\theta_{a,n}-\tilde\theta_{a,n+1})
\nonumber\\
&\quad
-
\sum_n
K_b
\sqrt{J_{b,n}J_{b,n+1}}
\cos(\tilde\theta_{b,n}-\tilde\theta_{b,n+1})
\nonumber\\
&\quad
-
g_{ab}
\sum_n
\sqrt{J_{a,n}J_{b,n}}
\cos(\tilde\theta_{a,n}+\tilde\theta_{b,n}).
\label{eq:SI_H_DTQC}
\end{align}
The last term pins the slow sum phase
\begin{equation}
\Theta_{+,n}
=
\tilde\theta_{a,n}+\tilde\theta_{b,n},
\label{eq:SI_DTQC_sum_phase}
\end{equation}
while the complementary phase
\begin{equation}
\Theta_{-,n}
=
\tilde\theta_{a,n}-\tilde\theta_{b,n}
\label{eq:SI_DTQC_difference_phase}
\end{equation}
is not directly pinned by the pump. Consequently, above threshold and when the winding is incommensurate with the drive, the local nonlinear attractor is a two-frequency torus. In the coupled lattice, spatial stiffness synchronizes the winding complementary phase across sites, establishing collective quasiperiodic motion under a single imposed clock.

\subsection{Frozen-action phase reduction and deterministic sum-locking tongue}
\label{subsec:SI_DTQC_phase_reduction}

We derive the deterministic sum-locking tongue by freezing the branch actions at their finite saturated values.

We first suppress the site index and retain the local part of Eq.~\eqref{eq:SI_H_DTQC},
\begin{equation}
\bar{\mathcal H}_{\rm loc}
=
\Delta_aJ_a+
\Delta_bJ_b+
\frac{\Lambda_{aa}}{2}J_a^2+
\frac{\Lambda_{bb}}{2}J_b^2+
\Lambda_{ab}J_aJ_b
-
g_{ab}\sqrt{J_aJ_b}\cos\Theta_+,
\qquad
\Theta_+=\tilde\theta_a+\tilde\theta_b .
\label{eq:SI_DTQC_local_H_phase_reduction}
\end{equation}
This expression assumes the nondegenerate sum-resonant sector, so the intrabranch \(2{:}1\) locking terms are off resonant or small compared with the retained sum-resonant channel. Uniform contributions from the spatial stiffness are absorbed into the renormalized rotating-frame detunings \(\Delta_\alpha\).

Hamilton's equations give
\begin{align}
\dot{\tilde\theta}_a
&=
\Delta_a+
\Lambda_{aa}J_a+
\Lambda_{ab}J_b
-
\frac{g_{ab}}{2}
\sqrt{\frac{J_b}{J_a}}
\cos\Theta_+,
\nonumber\\
\dot{\tilde\theta}_b
&=
\Delta_b+
\Lambda_{bb}J_b+
\Lambda_{ab}J_a
-
\frac{g_{ab}}{2}
\sqrt{\frac{J_a}{J_b}}
\cos\Theta_+ .
\label{eq:SI_DTQC_theta_dot_local}
\end{align}
Adding the two equations yields
\begin{equation}
\dot\Theta_+
=
\Delta_a+
\Delta_b+
\Lambda_{aa}J_a+
\Lambda_{bb}J_b+
\Lambda_{ab}(J_a+J_b)
-
\frac{g_{ab}}{2}
\left(
\sqrt{\frac{J_b}{J_a}}
+
\sqrt{\frac{J_a}{J_b}}
\right)
\cos\Theta_+ .
\label{eq:SI_DTQC_Theta_plus_full_local}
\end{equation}
For a spatially coherent saturated state with finite actions \(J_a=\bar J_a>0\) and \(J_b=\bar J_b>0\), this becomes the frozen-action Adler-type equation
\begin{equation}
\dot\Theta_+
=
\Delta_+^{\rm eff}
-
W_+\cos(\Theta_+-\Theta_0)
+
\cdots,
\label{eq:SI_DTQC_Adler}
\end{equation}
with
\begin{equation}
\Delta_+^{\rm eff}
=
\Delta_a+
\Delta_b+
\Lambda_{aa}\bar J_a+
\Lambda_{bb}\bar J_b+
\Lambda_{ab}(\bar J_a+\bar J_b),
\label{eq:SI_DTQC_Delta_eff}
\end{equation}
and
\begin{equation}
W_+
=
\frac{|g_{ab}|}{2}
\left(
\sqrt{\frac{\bar J_b}{\bar J_a}}
+
\sqrt{\frac{\bar J_a}{\bar J_b}}
\right).
\label{eq:SI_DTQC_W_plus}
\end{equation}
The phase offset \(\Theta_0\) incorporates the pump-coupling convention, and \(W_+\) is the sum-phase locking bandwidth. The local restoring rate near a stable fixed point scales as
\begin{equation}
\lambda_+
\sim
\sqrt{W_+^2-(\Delta_+^{\rm eff})^2},
\label{eq:SI_DTQC_restoring_rate}
\end{equation}
up to dissipative and saturation-dependent renormalizations.

A deterministic sum-locked fixed point of Eq.~\eqref{eq:SI_DTQC_Adler} can exist only when
\begin{equation}
|\Delta_+^{\rm eff}|\le W_+ .
\label{eq:SI_DTQC_sum_lock_condition}
\end{equation}

\paragraph*{Complementary phase and response frequencies.}
Subtracting the two lines of Eq.~\eqref{eq:SI_DTQC_theta_dot_local} gives
\begin{equation}
\dot\Theta_-
=
\Delta_a-\Delta_b
+\Lambda_{aa}\bar J_a-\Lambda_{bb}\bar J_b
+\Lambda_{ab}(\bar J_b-\bar J_a)
-\frac{g_{ab}}{2}
\left(
\sqrt{\frac{\bar J_b}{\bar J_a}}
-
\sqrt{\frac{\bar J_a}{\bar J_b}}
\right)\cos\Theta_+ ,
\label{eq:SI_DTQC_Theta_minus}
\end{equation}
which contains no \(\Theta_-\). With the uniform hopping shift \(-K_\alpha\), which Eq.~\eqref{eq:SI_DTQC_Delta_eff} keeps inside the renormalized \(\Delta_\alpha\), written explicitly, the sum of the two lines is the upper sign of Eq.~\eqref{eq:dtqc_phase_reduction} of the main text, with \(d_a+d_b=\Delta_+^{\rm eff}\), and Eq.~\eqref{eq:SI_DTQC_Theta_minus} is the lower sign. At a sum-locked fixed point \(\cos\Theta_+=(d_a+d_b)/[(g_{ab}/2)(\rho+\rho^{-1})]\), and Eq.~\eqref{eq:SI_DTQC_theta_dot_local} gives \(\dot{\tilde\theta}_a=-\dot{\tilde\theta}_b=(\bar J_ad_a-\bar J_bd_b)/(\bar J_a+\bar J_b)\), which is Eq.~\eqref{eq:dtqc_frequencies}. The Hamiltonian flow conserves \(\sum_n(J_{a,n}-J_{b,n})\), and in the uniform reduction \(J_a-J_b\) is conserved. For the equal branch damping used here the saturated actions are therefore equal, and the two lines sit at \(\Omega/2\pm(d_a-d_b)/2\). Since \(\Lambda>0\) and the lower branch is bus-like, the Kerr contribution to \(d_a-d_b\) is negative: the lower line moves down and the upper line moves up as the actions grow. At the operating modulation depth the second-order pump-induced shifts of the branch frequencies are comparable to the Kerr shifts, and the detunings in \(d_\alpha\) are to be read as including them.

\paragraph*{Relative-phase stiffness.}
Writing \(\tilde\theta_{a,n}=(\Theta_{+,n}+\Theta_{-,n})/2\) and \(\tilde\theta_{b,n}=(\Theta_{+,n}-\Theta_{-,n})/2\) in the hopping terms of Eq.~\eqref{eq:SI_H_DTQC}, with uniform actions and a uniform locked \(\Theta_+\), gives
\begin{equation}
\bar{\mathcal H}_{\rm hop}
=
-(K_a\bar J_a+K_b\bar J_b)
\sum_{n=1}^{N-1}
\cos\!\left(\frac{\Theta_{-,n+1}-\Theta_{-,n}}{2}\right),
\label{eq:SI_DTQC_hop}
\end{equation}
whose expansion to second order in the phase differences is Eq.~\eqref{eq:dtqc_stiffness}. With the bulk on-site detunings used in Eq.~\eqref{eq:slow_hamiltonian}, the free-end Laplacian of the circuit requires the boundary term
\begin{equation}
\bar{\mathcal H}_{\rm end}
=
-\frac{1}{2}\sum_{\alpha=a,b}K_\alpha\left(J_{\alpha,1}+J_{\alpha,N}\right),
\label{eq:SI_DTQC_end}
\end{equation}
which makes the uniform-state hopping shift equal to \(-K_\alpha\) at every site. Both hybrid bands of the circuit of Table~\ref{tab:circuit_bare_params} rise monotonically from \(k=0\), so \(K_a,K_b>0\) and the stiffness is positive. The uniform zero-twist state is the reference state because, at the DTQC operating point and for the damping of Table~\ref{tab:circuit_bare_params}, the linear Floquet growth rate exceeds the loss only for the \(k=0\) mode among the modes allowed in the \(N=12\) chain with the free-end boundary conditions of the circuit. Longer chains admit unstable modes with \(k\neq0\), and the reduction must then be taken about the corresponding state.

\subsection{Slow-flow equations and noise projection}
\label{subsec:SI_branch_slow_flow}

The deterministic branch-reduced dynamics follows from Hamilton's equations,
\begin{equation}
\dot J_{\alpha,n}
=
-\frac{\partial\bar{\mathcal H}}{\partial\tilde\theta_{\alpha,n}},
\qquad
\dot{\tilde\theta}_{\alpha,n}
=
\frac{\partial\bar{\mathcal H}}{\partial J_{\alpha,n}},
\qquad
\alpha\in\{a,b\}.
\label{eq:SI_slowflow_Hamilton_eqs}
\end{equation}
Dissipation and noise modify the action--angle dynamics in Eq.~\eqref{eq:SI_slowflow_Hamilton_eqs}. Within the same rotating-wave approximation as the slow Hamiltonian, let \(d\zeta_{\alpha,n}\) denote the projected circular complex noise, with
\begin{equation}
\langle d\zeta_{\alpha,n}d\zeta_{\beta,m}^{*}\rangle
=D_{\alpha\beta}\delta_{nm}\,dt,
\qquad
\langle d\zeta_{\alpha,n}d\zeta_{\beta,m}\rangle=0.
\label{eq:SI_slow_noise_covariance}
\end{equation}
For the canonical carrier normalization and rotating-wave noise convention used here, the covariance matrix follows from the hybrid rotation \(\mathsf R\) in Eq.~\eqref{eq:SI_circuit_hybrid_rotation}:
\begin{equation}
\mathsf D
=\mathsf R^{\mathsf T}
\begin{pmatrix}
\dfrac{D_Q}{C_0'\omega_{\mathrm{wg}}}&0\\[4pt]
0&\dfrac{D_Q}{C_r'\omega_r}
\end{pmatrix}
\mathsf R.
\label{eq:SI_projected_noise_matrix}
\end{equation}
Thus the diagonal and interbranch noise covariances are fixed by the input charge-noise intensity and the carrier projection, without introducing additional fitting parameters.

Using the previously defined amplitude \(c_{\alpha,n}=\sqrt{J_{\alpha,n}}e^{-i\tilde\theta_{\alpha,n}}\), It\^o's product rule gives
\begin{equation}
dJ_{\alpha,n}
=c_{\alpha,n}^{*}dc_{\alpha,n}
+c_{\alpha,n}dc_{\alpha,n}^{*}
+dc_{\alpha,n}^{*}dc_{\alpha,n},
\qquad
dc_{\alpha,n}^{*}dc_{\alpha,n}=D_{\alpha\alpha}\,dt.
\label{eq:SI_action_noise_drift}
\end{equation}
The resulting stochastic action--angle equations are
\begin{align}
dJ_{\alpha,n}
&=\left[-\frac{\partial\bar{\mathcal H}}{\partial\tilde\theta_{\alpha,n}}-\Gamma_\alpha^{(J)}J_{\alpha,n}+D_{\alpha\alpha}\right]dt
+2\operatorname{Re}\!\left(c_{\alpha,n}^{*}d\zeta_{\alpha,n}\right),
\nonumber\\
d\tilde\theta_{\alpha,n}
&=\frac{\partial\bar{\mathcal H}}{\partial J_{\alpha,n}}\,dt
-\operatorname{Im}\!\left(\frac{d\zeta_{\alpha,n}}{c_{\alpha,n}}\right).
\label{eq:SI_slowflow_damped_noisy}
\end{align}
For the present equal-damping parameters, \(\Gamma_a^{(J)}=\Gamma_b^{(J)}=R_0/L_s=R_r/L_r\). The circular-noise approximation gives no additional It\^o phase drift for \(J_{\alpha,n}>0\).

The positive drift \(D_{\alpha\alpha}\) is fixed by the projected noise covariance and vanishes with the input noise strength. The driven stationary actions are determined dynamically by pumping, nonlinearity, and dissipation. The action and phase noise terms inherit the same projected noise channels, including any interbranch correlations.

The activation analysis in Figs.~2(d,g) and 3(g) uses stochastic trajectories of Eq.~\eqref{eq:SI_slowflow_damped_noisy} with charge-noise intensity $D_Q$, initialized in the ordered state. Escape events are identified from a sustained loss of the ordered-state overlap, evaluated from the subharmonic response in the DTC sector and the two-component sum-locked response in the DTQC sector. For short arrays, these events coincide with the loss of order across the chain. For longer arrays, they capture local nucleation preceding the spatial propagation of the escape.

The escape rate \(\Gamma\) is the number of escapes divided by the total time at risk. For each array size \(N\), the effective activation coefficient is extracted using a Poisson-likelihood fit to the Arrhenius form
\begin{equation}
\ln(\Gamma)=a_N-\frac{E_A(N)}{D_Q},
\end{equation}
where \(\Gamma\) is expressed in \(\mathrm{s^{-1}}\), \(a_N\) is the logarithmic prefactor, and \(E_A(N)\), with units of \(\mathrm{C^2/s}\), characterizes the noise dependence of the escape rate. Both coefficients are fitted independently for each \(N\) over the noise range exhibiting Arrhenius behavior.

The phase resets in Fig.~3(d) are read from the continuously lifted sum phase. At $\delta_{\rm eff}=0.125$ the sum phase stays locked over the simulated $5\times10^5T$. At $\delta_{\rm eff}=0.113$ the trajectories show locked plateaus separated by noise-activated transitions, and successive plateaus differ by integer multiples of $2\pi$. During a transition the response amplitude collapses before the sum phase relocks, so that an escape is a transient loss and recovery of the ordered response.

The hysteretic onset in Fig.~3(e) is examined by comparing stochastic and noise-free modulation sweeps. The noise-free dynamics converges to the same stationary branch for increasing and decreasing modulation depths, whereas finite charge noise produces hysteretic separation between the sweeps. This behavior reflects noise-activated switching and finite-time persistence of the ordered response during parameter variation.

\subsection{Validation of the rotating-wave reduction}
\label{subsec:SI_branch_validation}

We compare the rotating-wave-averaged Hamiltonian with direct time-domain integration of the reduced circuit Hamiltonian in Eq.~\eqref{eq:SI_circuit_H_full_reduced}, prior to rotating-wave averaging. The three resonance channels are examined at $\Omega/2\pi=2.267$~GHz and $\delta_{\rm eff}=0.589$ for DTC$_a$, $2.98$~GHz and $0.06$ for DTC$_b$, and $2.47$~GHz and $0.12$ for DTQC. The DTC$_a$ drive selects the lower band near $kd/\pi\simeq0.5$. The branch actions, slow phases, and response spectra obtained from the two descriptions agree across the degenerate and nondegenerate resonance channels (Fig.~\ref{fig:SI_AA_vs_exact_branch_validation}).

\begin{figure*}[htb]
  \centering
  \includegraphics[width=1.00\textwidth]{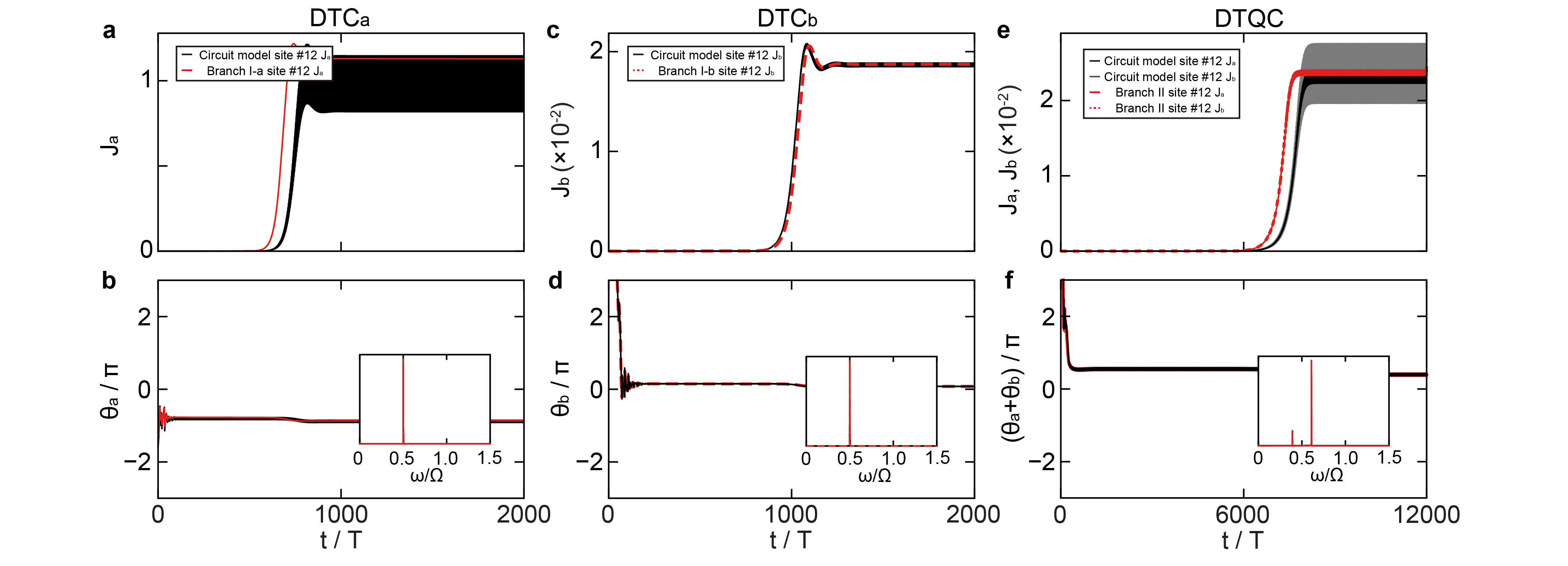}
  \caption{\textbf{Validation of the rotating-wave reduction.} Black traces are obtained from direct integration of the unaveraged reduced circuit Hamiltonian, while red traces show the corresponding rotating-wave-averaged dynamics. Both calculations use identical circuit parameters and drive conditions. The time axis is normalized by the drive period \(T=2\pi/\Omega\).
  (a,b) DTC\(_a\) (\(\Omega/2\pi=2.267\) GHz, \(\delta_{\rm eff}=0.589\)): action \(J_a\), including its within-period oscillation, and slow phase \(\tilde\theta_a\).
  (c,d) DTC\(_b\) (\(2.98\) GHz, \(\delta_{\rm eff}=0.06\)): action \(J_b\) and slow phase \(\tilde\theta_b\); slow-flow results dashed.
  (e,f) DTQC (\(2.47\) GHz, \(\delta_{\rm eff}=0.12\)): actions \(J_a\) (black, red solid) and \(J_b\) (grey, red dashed), and the slow sum phase \((\tilde\theta_a+\tilde\theta_b)/\pi\).
  Insets show normalized spectra of the probe signal: a single subharmonic line for the DTC channels and two lines for the DTQC channel.}
  \label{fig:SI_AA_vs_exact_branch_validation}
\end{figure*}

Branch I consists of the two degenerate parametric resonance channels DTC$_a$ and DTC$_b$, realized near $\Omega\simeq2\omega_a$ and $\Omega\simeq2\omega_b$, respectively. Branch II is the nondegenerate sum-resonant DTQC channel, realized near $\Omega\simeq\omega_a+\omega_b$. In the weak-mixing limit, these hybrid-branch conditions reduce continuously to the corresponding bare-sector conditions.

\section{Experimental Implementation}
\label{sec:ExperimentalImplementation}
\subsection{Experimental platform, measurement setup, and noise injection}
To implement the unit cell of the nonlinear LC resonator, a metallic split-ring resonator (SRR) was patterned onto a printed circuit board (Fig.~\ref{fig:Figsm}b). Each unit cell incorporates a varactor diode (SMV1247, Skyworks), whose capacitance can be modulated by applying a DC-biased AC voltage across it. As a result, the resonant characteristics of the unit cell undergo time-periodic modulation. In the experiment, a DC bias voltage of 2.6~V is applied. The discrete time-crystalline (DTC) phase is observed at $\Omega=2\pi f_{\mathrm{mod}}$ with $f_{\mathrm{mod}}=2.98~\mathrm{GHz}$, while the discrete time-quasicrystalline (DTQC) phase is observed at $\Omega=2\pi f_{\mathrm{mod}}$ with $f_{\mathrm{mod}}=2.47~\mathrm{GHz}$. 

To construct a one-dimensional array of coupled nonlinear LC resonators, 12 unit cells were placed inside a waveguide (Fig.~\ref{fig:Figsm}a). The complete experimental setup is shown in Fig.~\ref{fig:Figsm}c. The driving signal (AC voltage) is generated by a signal generator (N5171B, Keysight) and amplified by a radio-frequency amplifier (RUM43020-10, RFHIC). The amplified AC voltage is then DC-biased using a bias tee and subsequently distributed through a two-way splitter, followed by a six-way splitter. Additional two-way splitters are inserted just before the resonator array, with each splitter enabling individual signal measurement from a corresponding unit cell. To prevent unwanted DC current flow, a DC block is inserted at one end of the two-way splitter. Additionally, a low-pass filter (LPF) is placed before the oscilloscope (DPO 70804, Tektronix) or spectrum analyzer (E4404B, Agilent) to suppress reflections of the driving signal. The radio-frequency modulation distributed by the splitter network is the only coherent external temporal drive. 
Fixed cable delays and splitter phases correspond to static phase offsets of the same global tone and do not introduce an additional modulation frequency.

Noise is injected from one end of the waveguide using a noise source module (PE85N1012, Pasternack), which generates broadband noise in the 10~MHz to 3~GHz range. This noise is further amplified by a high-power amplifier (ZHL-10W-202-S+, Mini-Circuits). A variable attenuator (PE7422, Pasternack) is used to control the noise intensity injected into the system. Fig.~\ref{fig:Figsm2} shows the power spectrum of the noise source measured before entering the waveguide, with a 30~dB variable attenuator applied. The amplifier operates over a frequency range of 10~MHz to 2~GHz, within which the power level is approximately --93.34~dBm/Hz. This value is used in the main text as a reference for characterizing the noise power spectral density. The injected noise is broadband and phase-incoherent and therefore carries no second clock. It seeds fluctuations and controls how sharply the ordered response sets in, while the response frequencies and the locked sum phase are selected by the driven nonlinear lattice.

\begin{figure*}[htb]
  \centering
  \includegraphics[width=0.9\textwidth]{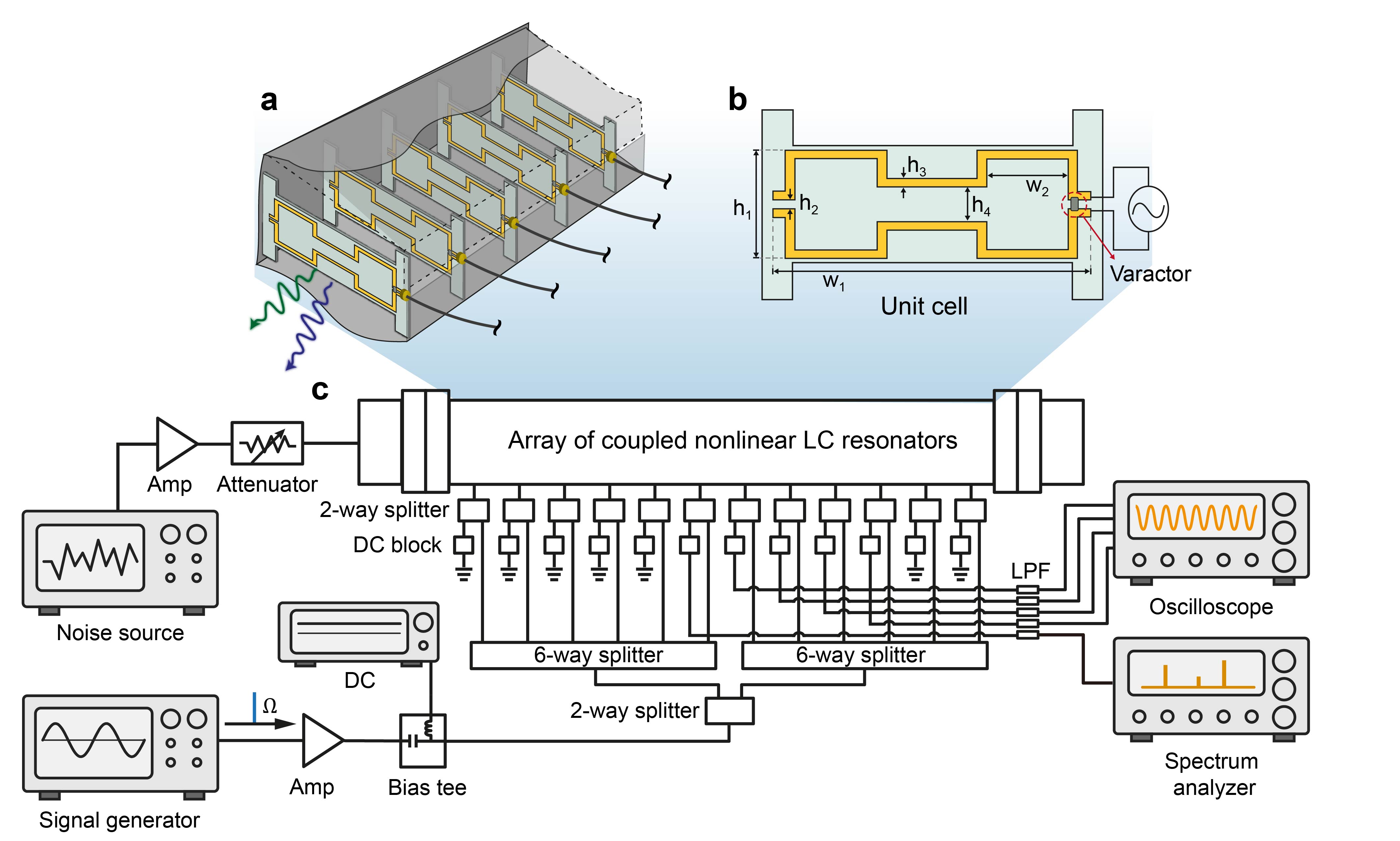}
  \caption{\textbf{Experimental platform: modulated nonlinear LC-resonator array in a waveguide.}
  (a) Schematic of a one-dimensional array of $N=12$ coupled nonlinear LC resonators embedded in a waveguide. (b) Geometry of a single unit cell incorporating a varactor diode that enables time-periodic modulation of the capacitance via an AC drive voltage applied on top of a DC bias. The unit-cell dimensions are   $h_1=40~\mathrm{mm}$, $h_2=1~\mathrm{mm}$, $h_3=3~\mathrm{mm}$, $h_4=10~\mathrm{mm}$, $w_1=105~\mathrm{mm}$, and $w_2=27.5~\mathrm{mm}$.   (c) Block diagram of the experimental measurement setup used to drive the modulation and record the waveguide response.}
  \label{fig:Figsm}
\end{figure*}

\begin{figure*}[htb]
  \centering
  \includegraphics[width=0.9\textwidth]{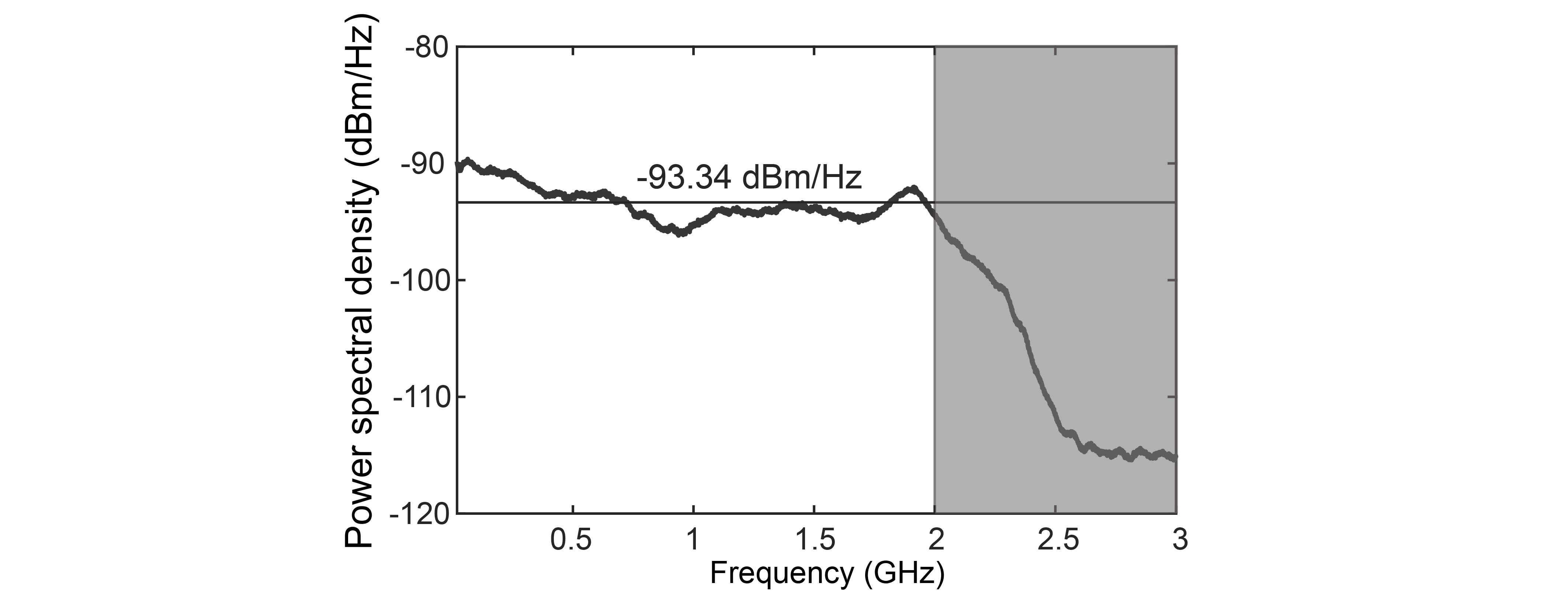}
  \caption{\textbf{Characterization of the injected noise spectrum.}
  Power spectral density of the external noise source measured before injection into the waveguide, using a $30~\mathrm{dB}$ attenuator. The mean level over the $10~\mathrm{MHz}$--$2~\mathrm{GHz}$ band is approximately $-93.34~\mathrm{dBm/Hz}$, which is used as the reference noise power spectral density in the experiments.}
  \label{fig:Figsm2}
\end{figure*}

\subsection{Signal processing and reconstruction of experimental DTQC observables}
\label{subsec:SI_DTQC_reconstruction}

The experimental DTQC observables in Fig.~\ref{fig:Fig4}(i--k) and Supplementary Fig.~\ref{fig:Figsm_spatial_coherence}(a--c) are reconstructed from the measured cell-resolved voltage traces \(V_n(t)\). The measured-site set is the simultaneously recorded four-channel set \(\mathcal M=\{7,8,9,10\}\), with \(N_{\mathcal M}=|\mathcal M|=4\), and the representative-site phase reconstruction in Fig.~\ref{fig:Fig4}(k) is shown for \(n_0=7\). The DTQC data are analyzed at the modulation frequency \(f_{\mathrm{mod}}=\Omega/2\pi=2.47~\mathrm{GHz}\), using the original oscilloscope time grid at \(25~\mathrm{GSa/s}\).

For each measured site, the two response peaks are identified within fixed lower and upper frequency windows, using the same extraction procedure throughout the experimental scan.

For the experimental phase diagram, the two-line spectral weight is evaluated from the two peak heights
\begin{equation}
M_{\mathrm{DTQC}}^{\mathrm{spec}}
=
\left(A_1^{\mathrm{pk}}A_2^{\mathrm{pk}}\right)^{1/2}.
\label{eq:SI_MDTQC_spec_definition}
\end{equation}
Here \(A_j^{\mathrm{pk}}\) denotes the linear spectral peak power in the fixed frequency window assigned to the \(j\)-th response component, and \(P_j^{\mathrm{pk}}[\mathrm{dBm}]\) denotes its logarithmic value. The same peak windows and extraction procedure are used throughout the \((S_{\mathrm{noise}},P_d)\) scan, so that the two-line spectral weight tracks the joint appearance of the two response peaks.

When the spectra are displayed in logarithmic power units, the displayed two-line spectral weight is written as
\begin{equation}
M_{\mathrm{DTQC}}^{\mathrm{spec}}[\mathrm{dBm}]
=
\frac{
P_1^{\mathrm{pk}}[\mathrm{dBm}]
+
P_2^{\mathrm{pk}}[\mathrm{dBm}]
}{2},
\label{eq:SI_MDTQC_dBm}
\end{equation}
which is the logarithmic representation of the geometric mean of the corresponding linear peak powers. The two response frequencies are also tracked independently across the experimental $(S_{\mathrm{noise}},P_d)$ scan. Both spectral peaks stand more than $25$~dB above the background at $126$ of the $184$ operating points, which span $P_d=-10.45$ to $-9.70$~dBm. Across these points the lower and upper frequencies move in opposite directions by about $0.1$~MHz (correlation coefficient $-0.95$), while their sum varies by $33$~kHz and lies $0.39$~MHz below the nominal modulation frequency, inside the $1$~MHz frequency sampling of the spectrum analyzer.

Throughout the phase reconstruction and quasiperiodicity analysis, we use angular frequencies. At the representative experimental operating point $P_d=-10.3$~dBm, marked in Fig.~\ref{fig:Fig4}(h), the selected response frequencies and imposed modulation frequency are
\begin{equation}
\frac{\omega_1}{2\pi}=0.968~\mathrm{GHz},\qquad \frac{\omega_2}{2\pi}=1.502~\mathrm{GHz},\qquad \frac{\Omega}{2\pi}=2.47~\mathrm{GHz}.
\label{eq:SI_selected_omega_pair}
\end{equation}
The selected pair \((\omega_1,\omega_2)\) defines the two narrow-band voltage components used for phase extraction and for the quasiperiodicity tests in Fig.~\ref{fig:Fig5}. The numerical values in Eq.~\eqref{eq:SI_selected_omega_pair} are rounded for display; the phase reconstruction and the low-order relation map use the extracted frequencies before rounding.

The independent frequency analysis uses voltage records acquired simultaneously at sites $7$--$10$ for $1~\mu\mathrm{s}$ at $25$~GSa/s, containing $25\,000$ samples per channel and a nominal Fourier resolution of $1$~MHz. The lower and upper peaks are determined separately within $0.80$--$1.20$~GHz and $1.30$--$1.70$~GHz using Hann-windowed spectra, zero-padding to $2^{20}$ points, and parabolic peak interpolation.

In the independent phase-reconstruction analysis, each response component is isolated using a zero-phase fourth-order Butterworth band-pass filter with a $50$~MHz full bandwidth centered at its independently extracted peak. The instantaneous phases are reconstructed from the Hilbert analytic signals after discarding $150$ drive periods at each end of the record. Linear fits to the unwrapped phases give $\omega_1/2\pi=0.967905$~GHz and $\omega_2/2\pi=1.502096$~GHz. The sum-frequency residual across the four measured sites is $0.6$--$2.9$~kHz, corresponding to $|\omega_1+\omega_2-\Omega|/\Omega \leq1.2\times10^{-6}$, and remains below $6\times10^{-6}$ in seven overlapping sub-windows.

Varying the filter bandwidth from $20$ to $100$~MHz and its order between $2$ and $4$ changes the sum-frequency residual by less than $0.4$~kHz. At all four sites the temporal circular concentration $R_{\Theta_+,n}\equiv|\langle e^{i\Theta_{+,n}(t)}\rangle_t|$ exceeds $0.9999$, the response envelopes stay above $0.91$ of their median amplitudes, and the sum-phase fluctuations have standard deviations of about $0.003\pi$.

From the two corresponding narrow-band voltage components, we construct analytic signals
\begin{equation}
\mathcal A_n^{(j)}(t)
=
V_n^{(j)}(t)
+i\,\mathsf H\!\left[V_n^{(j)}(t)\right],
\qquad j=1,2,
\label{eq:SI_analytic_signal}
\end{equation}
where $\mathsf H$ denotes the Hilbert transform. We write
\begin{equation}
\mathcal A_n^{(j)}(t)
=
R_{j,n}(t)e^{i\phi_{j,n}(t)},
\qquad
R_{j,n}(t)>0,
\label{eq:SI_analytic_signal_phase}
\end{equation}
with \(\phi_{j,n}(t)\) chosen continuously over the post-transient analysis window. 
To use the same rotating-frame convention as the branch-reduced normal form, we define
\begin{equation}
\tilde{\phi}_{j,n}(t)
=
\phi_{j,n}(t)-\frac{\Omega t}{2},
\qquad j=1,2 .
\label{eq:SI_slow_phase_def}
\end{equation}
For the representative site \(n_0=7\), the reconstructed sum phase is defined modulo $2\pi$ as
\begin{equation}
\Theta_{+,n_0}(t) = \tilde{\phi}_{1,n_0}(t)+\tilde{\phi}_{2,n_0}(t) \quad (\mathrm{mod}\ 2\pi),
\label{eq:SI_exp_sum_phase}
\end{equation}
and the complementary phase is kept as a continuous winding coordinate,
\begin{equation}
\Theta_{-,n_0}(t) = \tilde{\phi}_{1,n_0}(t)-\tilde{\phi}_{2,n_0}(t).
\label{eq:SI_exp_difference_phase}
\end{equation}

Equivalently,
\begin{equation}
\Theta_{+,n_0}(t) = \phi_{1,n_0}(t)+\phi_{2,n_0}(t)-\Omega t \quad (\mathrm{mod}\ 2\pi),
\qquad
\Theta_{-,n_0}(t)
=
\phi_{1,n_0}(t)-\phi_{2,n_0}(t).
\label{eq:SI_exp_sum_difference_phase_lab}
\end{equation}
The sum phase \(\Theta_{+,n_0}\) measures locking of the two-frequency response to the imposed pump, whereas \(\Theta_{-,n_0}\) is the winding phase of the unpinned complementary degree of freedom. Figure~\ref{fig:Fig4}(k) is obtained by plotting these two quantities for the representative-site voltage trace used in Fig.~\ref{fig:Fig4}(i,j).

The two-phase trajectory shown in Fig.~\ref{fig:Fig5}(c) and the recurrence measure in Fig.~\ref{fig:Fig5}(d) are reconstructed from the same representative-site signal using the same half-pump rotating phase coordinates \((\tilde\phi_{1,n_0},\tilde\phi_{2,n_0})\). Their quantitative definitions are given in Sec.~\ref{subsec:SI_DTQC_quasiperiodicity_diagnostics}.

To quantify the phase-referenced sum-phase memory used in the subset analysis of Supplementary Fig.~\ref{fig:Figsm_spatial_coherence}(c), we specialize the main-text complex phase memory to the measured-site set,
\begin{equation}
Z_{\Theta_+}^{(\mathcal M)}(t)
\equiv
\frac{1}{N_{\mathcal M}}
\sum_{n\in\mathcal M}
e^{i\Delta\Theta_{+,n}(t)},
\qquad
\Delta\Theta_{+,n}(t)
=
\Theta_{+,n}(t)-\Theta_{+,n}(t_{\rm ref})
\quad (\mathrm{mod}\ 2\pi).
\label{eq:SI_ZTheta_measured}
\end{equation}
Here $t_{\rm ref}$ is the first time point of the post-transient analysis window. The site-dependent phase reference removes static channel offsets associated with cable and probe delays. For each subset $\mathcal S\subseteq\mathcal M$, the same definition applies with $\mathcal M$ replaced by $\mathcal S$. The subset phase memories are compared with finite-site random-phase references matched to $N_{\mathcal S}=3$, as shown in Supplementary Fig.~\ref{fig:Figsm_spatial_coherence}(c).


\subsection{Quantitative quasiperiodicity tests for the experimental DTQC state}
\label{subsec:SI_DTQC_quasiperiodicity_diagnostics}

The quasiperiodicity tests in Fig.~\ref{fig:Fig5} are evaluated at the same representative operating point and from the same selected angular-frequency pair \((\omega_1,\omega_2)\) used for the phase reconstruction and measured-site coherence analysis in Sec.~\ref{subsec:SI_DTQC_reconstruction}. They distinguish a quasiperiodic DTQC state from low-order locking to the single imposed modulation.

We first test small-integer relations among the two response frequencies and the imposed modulation frequency. For each integer pair \((l_1,l_2)\), we define the normalized low-order mismatch
\begin{equation}
\Delta_{l_1,l_2}
=
\min_{l_3\in\mathbb Z}
\frac{
\left|
l_1\omega_1+l_2\omega_2-l_3\Omega
\right|
}{\Omega}.
\label{eq:SI_low_order_mismatch}
\end{equation}
The relation map in Fig.~\ref{fig:Fig5}(a) is evaluated using the independently extracted response frequencies before rounding. The fundamental sum-resonance relation exhibits a normalized mismatch of order $10^{-6}$, with its integer multiples forming the low-mismatch diagonal.

Equivalently, \(l_3\) is chosen as the nearest integer to \((l_1\omega_1+l_2\omega_2)/\Omega\). In the relation map shown in Fig.~\ref{fig:Fig5}(a), the integer pair \((l_1,l_2)\) is restricted to a finite low-order window, excluding the trivial zero vector. For Fig.~\ref{fig:Fig5}(a), we use \(|l_1|,|l_2|\leq L_{\max}\) with \(L_{\max}=8\). The low-mismatch diagonal corresponds to integer multiples of the expected sum-resonant condition
\begin{equation}
\omega_1+\omega_2\simeq\Omega,
\label{eq:SI_expected_sum_resonance}
\end{equation}
whereas additional low-order frequency locking would appear as other isolated low-mismatch points away from this sum-resonant family. Within the experimental frequency resolution, no such additional low-order relation is observed.

The combination-line overlay in Fig.~\ref{fig:Fig5}(b) is constructed from the same two extracted angular frequencies. Candidate spectral lines are generated as
\begin{equation}
\omega_{m,n}
=
m\omega_1+n\omega_2,
\qquad
m,n\in\mathbb Z,
\label{eq:SI_combination_lines}
\end{equation}
and only positive frequencies inside the plotted spectral window are retained. In practice, the displayed set is restricted to low-order combinations. For Fig.~\ref{fig:Fig5}(b), we retain combinations satisfying \(|m|+|n|\leq M_{\max}\) with \(M_{\max}=4\). The construction uses no fitting frequencies beyond the two fundamentals of the phase reconstruction.

For the two-phase reconstruction in Fig.~\ref{fig:Fig5}(c), we use the half-pump rotating phases modulo \(2\pi\),
\begin{equation}
\left(
\tilde\phi_{1,n_0}(t)\ {\rm mod}\ 2\pi,\,
\tilde\phi_{2,n_0}(t)\ {\rm mod}\ 2\pi
\right)
\in \mathbb T^2,
\label{eq:SI_two_phase_coordinates}
\end{equation}
where \(\tilde\phi_{j,n}\) is defined in Eq.~\eqref{eq:SI_slow_phase_def}. 
The reconstructed trajectory provides a signal-level projection of the measured voltage response onto the two selected DTQC components in the half-pump rotating frame. The same coordinates are used for the recurrence measure in Fig.~\ref{fig:Fig5}(d).

For positive integer \(m\), we evaluate recurrence at lags \(\tau=2mT\), for which the half-pump rotating and laboratory-frame phase increments coincide modulo \(2\pi\). The normalized minimum return distance is
\begin{equation}
d_{\min}(\tau)
=
\min_{t\in\mathcal W_\tau}
\frac{1}{\sqrt{2}\pi}
\left[
d_{\mathbb T}^2
\!\left(
\tilde\phi_{1,n_0}(t+\tau),
\tilde\phi_{1,n_0}(t)
\right)
+
d_{\mathbb T}^2
\!\left(
\tilde\phi_{2,n_0}(t+\tau),
\tilde\phi_{2,n_0}(t)
\right)
\right]^{1/2},
\label{eq:SI_min_return_distance}
\end{equation}
where \(\mathcal W_\tau\) is the post-transient analysis window for which both \(t\) and \(t+\tau\) lie inside the measured trace, and
\begin{equation}
d_{\mathbb T}(\phi,\phi') = \min_{m\in\mathbb Z} \left| \phi-\phi'+2\pi m \right|.
\label{eq:SI_torus_phase_distance}
\end{equation}
Here $d_{\mathbb T}$ denotes the wrapped angular separation on the circle. The normalization gives \(0\le d_{\min}\le1\). We assess these return distances together with the low-order frequency-relation map to characterize recurrence of the quasiperiodic response relative to the imposed pump.

The low-order integer-relation test, the two-frequency combination spectrum, the half-pump rotating two-phase reconstruction, and the recurrence analysis together support the interpretation of the observed response as a quasiperiodic DTQC state.

\subsection{Spatial coherence and collective synchronization of the DTQC state}
\label{subsec:SI_DTQC_spatial_coherence}

To characterize collective synchronization across the measured region, we evaluate pairwise and distance-resolved phase coherences on the simultaneously recorded site set $\mathcal M=\{7,8,9,10\}$, and test how the phase-referenced sum-phase memory depends on the choice of subset.

Using the reconstructed site-resolved phases, we define the pairwise sum- and complementary-phase coherences as
\begin{equation}
C_{ij}^{\pm}\equiv\left|\left\langle e^{i[\Theta_{\pm,i}(t)-\Theta_{\pm,j}(t)]}\right\rangle_t\right|.
\label{eq:SI_Cij_DTQC}
\end{equation}
The distance-resolved coherence and mean pairwise coherence are then given by
\begin{equation}
C_{\pm}(r)\equiv\left\langle C_{ij}^{\pm}\right\rangle_{|i-j|=r},\qquad C_{\pm}\equiv\frac{2}{N_{\mathcal M}(N_{\mathcal M}-1)}\sum_{\substack{i<j\\i,j\in\mathcal M}}C_{ij}^{\pm}.
\label{eq:SI_spatial_coherence_averages}
\end{equation}
Here $C_{ij}\equiv C_{ij}^{+}$ denotes the sum-phase correlator used in the subset analysis. To resolve the relative-phase dynamics, we introduce $\Delta\Theta_{-,ij}(t)=\operatorname{unwrap}[\Theta_{-,i}(t)-\Theta_{-,j}(t)]$ and subtract its temporal mean,
\begin{equation}
\delta\Theta_{-,ij}(t)=\Delta\Theta_{-,ij}(t)-\left\langle\Delta\Theta_{-,ij}(t)\right\rangle_t.
\label{eq:SI_relative_phase_drift}
\end{equation}
This removes the static pair-dependent offsets and isolates the temporal evolution of the relative complementary phase.

\begin{figure*}[htb]
  \centering
  \includegraphics[width=0.9\textwidth]{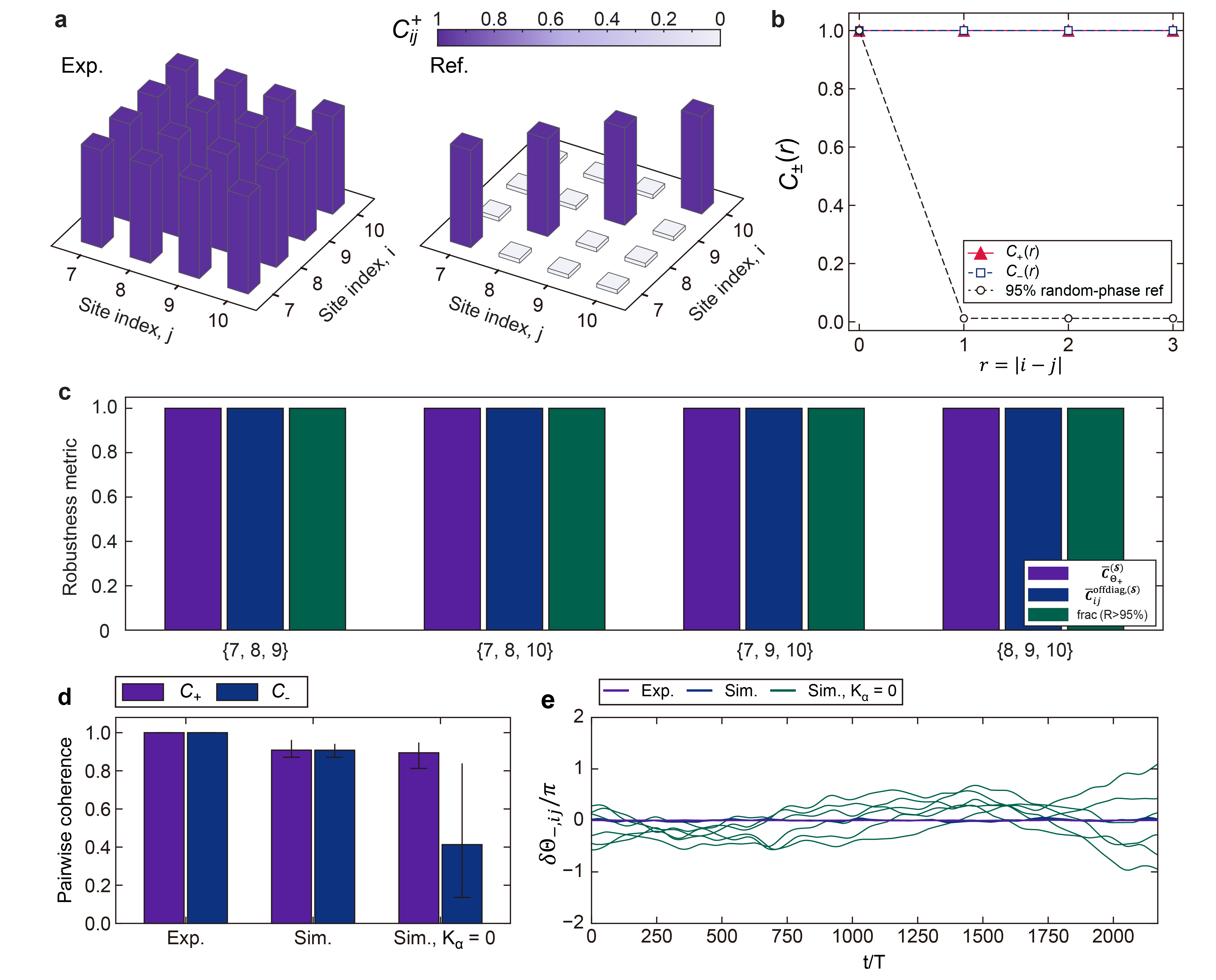}
    \caption{\textbf{Spatial coherence and collective synchronization of the DTQC state.} (a) Experimental pairwise sum-phase coherence $C_{ij}^{+}$ across sites $7$--$10$ and the corresponding 95\% random-phase reference. (b) Distance-resolved sum- and complementary-phase coherences $C_{\pm}(r)$ and their random-phase references. (c) Subset robustness of the phase-referenced sum-phase coherence over all three-site subsets, characterized by the time-averaged phase memory, mean pairwise coherence, and fraction exceeding the 95\% random-phase reference. (d) Mean pairwise coherences $C_+$ and $C_-$ in experiment, coupled simulation, and uncoupled simulation ($K_\alpha=0$). Whiskers indicate the minimum and maximum values across site pairs and noise realizations. (e) Mean-centered relative complementary-phase trajectories $\delta\Theta_{-,ij}(t)/\pi$, showing stable relative phases in the experiment and coupled simulation and progressive phase drift without inter-site coupling.}
  \label{fig:Figsm_spatial_coherence}
\end{figure*}

Figure~\ref{fig:Figsm_spatial_coherence}(a) shows the pairwise sum-phase coherence across the measured sites. The 95\% random-phase references for the pairwise coherences in panels (a,b) are obtained from independent random-phase time series matched to the analysis window and sample count, with the diagonal entries remaining unity. The finite-site random-phase reference for the subset phase memory in panel (c) is evaluated separately using the random-phasor distribution for $N_{\mathcal S}=3$.

The distance-resolved correlations in Fig.~\ref{fig:Figsm_spatial_coherence}(b) show that both the sum and complementary phases maintain near-unity coherence across all measured site separations, well above the random-phase references.

To assess spatial coherence across the measured region, we evaluate all three-site subsets of $\mathcal M=\{7,8,9,10\}$ using the phase-referenced sum-phase memory defined in Eq.~\eqref{eq:SI_ZTheta_measured}. For each subset $\mathcal S$, we evaluate the time-averaged sum-phase memory
\begin{equation}
\overline{\mathcal C}_{\Theta_+}^{(\mathcal S)}
\equiv
\left\langle
\left|Z_{\Theta_+}^{(\mathcal S)}(t)\right|
\right\rangle_t,
\label{eq:SI_Cbar_subset_main}
\end{equation}
the mean off-diagonal pairwise coherence
\begin{equation}
\overline{C}_{ij}^{\mathrm{offdiag},(\mathcal S)}
\equiv
\frac{2}{N_{\mathcal S}(N_{\mathcal S}-1)}
\sum_{\substack{i<j\\i,j\in\mathcal S}} C_{ij}^{+},
\label{eq:SI_Cbar_subset_pair}
\end{equation}
and the fraction of samples for which $|Z_{\Theta_+}^{(\mathcal S)}(t)|$ exceeds the 95\% random-phase reference calculated for $N_{\mathcal S}=3$. All three measures stay high for every subset (Fig.~\ref{fig:Figsm_spatial_coherence}(c)), so the spatial coherence does not depend on the chosen three-site combination.

The complementary-phase coherence provides a measure of collective synchronization, with the experiment exhibiting $C_-\simeq0.9999$. To examine its dependence on inter-site coupling, we compare the coupled lattice with an uncoupled control ($K_\alpha=0$). The simulations use $N=12$, $\delta_{\rm eff}=0.113$, and $D_Q=10^{-15.52}~\mathrm{C^2/s}$, with eight independent noise realizations and random initial conditions. The experimental and simulated coherences are evaluated over matched post-transient windows of $2170T$, using the same 50~MHz Butterworth--Hilbert phase-reconstruction procedure. Pairwise coherences are averaged over the six site pairs among sites $7$--$10$ and, for simulations, eight noise realizations. The experiment corresponds to $P_d=-10.3$~dBm.

The mean pairwise sum-phase coherence changes little upon removing inter-site coupling, whereas the complementary-phase coherence is substantially reduced (Fig.~\ref{fig:Figsm_spatial_coherence}(d)). The relative complementary-phase differences remain stable in the experiment and coupled simulation, while the uncoupled sites develop progressive phase drift (Fig.~\ref{fig:Figsm_spatial_coherence}(e)). These observations distinguish pump-induced sum-phase locking from interaction-mediated synchronization of the winding complementary phase.

\clearpage
\end{widetext}
\end{document}